\documentclass{aa}
\usepackage[varg]{txfonts}
\usepackage{multirow}
\usepackage{threeparttable}
\usepackage{natbib}

\defcitealias{Hill2019}{H19}
\defcitealias{Reichert2020}{R20}

\usepackage[normalem]{ulem}

\begin{document}


\title{Near-infrared chemical abundances of \\stars in the Sculptor dwarf galaxy}


 \author{Baitian Tang\inst{\ref{inst1},\ref{inst2}}\thanks{email: tangbt@mail.sysu.edu.cn} \and Jiajun~Zhang\inst{\ref{inst1},\ref{inst2}\thanks{email: zhangjj239@mail.sysu.edu.cn}} \and
 Zhiqiang Yan\inst{\ref{inst3},\ref{inst4}\thanks{email: yan@nju.edu.cn}} \and Zhiyu Zhang\inst{\ref{inst3},\ref{inst4}} \and 
 Leticia Carigi\inst{\ref{inst5}} \and Jos\'e G. Fern\'andez-Trincado\inst{\ref{inst6}} }
\institute{School of Physics and Astronomy, Sun Yat-sen University, Zhuhai 519082, China \label{inst1}
\and
CSST Science Center for the Guangdong–Hong Kong–Macau Greater Bay Area, Zhuhai, 519082, China\label{inst2}
\and
 School of Astronomy and Space Science, Nanjing University, Nanjing 210093, China\label{inst3}
\and
Key Laboratory of Modern Astronomy and Astrophysics (Nanjing University), Ministry of Education, Nanjing 210093, China\label{inst4}
\and 
Universidad Nacional Aut\'onoma de M\'exico, Instituto de Astronom\'ia, AP 70-264, CDMX  04510, M\'exico\label{inst5}
\and
Instituto de Astronom\'ia, Universidad Cat\'olica del Norte, Av. Angamos 0610, Antofagasta, Chile\label{inst6}
}

\date{Received   / Accepted  }

\abstract{
Owing to the recent identification of major substructures in our Milky Way (MW), the astronomical community has started to reevaluate the importance of dissolved and existing dwarf galaxies. In this work, we investigate up to 13 elements in 43 giant stars of the Sculptor dwarf galaxy (Scl) using high-signal-to-noise-ratio near-infrared (NIR) APOGEE spectra. Thanks to the strong feature lines in the NIR,  we were able to determine high-resolution O, Si, and Al abundances for a large group of sample stars for the first time in Scl. By comparing the [$\alpha$/Fe] (i.e., O, Mg, Si, Ca, and Ti) of the stars in Scl, Sagittarius, and the MW, we confirm the general trend that less massive galaxies tend to show lower [$\alpha$/Fe]. The low [Al/Fe] ($\sim -0.5$) in Scl demonstrates the value of this ratio as a discriminator with which to identify stars born in dwarf galaxies (from MW field stars). A chemical-evolution model suggests that Scl has a top-light initial mass function (IMF), with a high-mass IMF power index of $\sim -2.7$, and a minimum Type Ia supernovae delay time of $\sim 100$ Myr. Furthermore, a linear regression analysis indicates a negative radial metallicity gradient and positive radial gradients for [Mg/Fe] and [Ca/Fe], in qualitative agreement with the outside-in formation scenario. 
}


\maketitle

\section{Introduction}
\label{sec:introduction}

Under the hierarchical galaxy formation scenario, dwarf galaxies are first formed in dark matter subhalos, and later merge to form bigger galaxies, such as our Milky Way (MW). Some of the early mergers can be traced back using their clustering in orbital and kinematical parameter space, and one prominent example is the discovery of Gaia-Sausage-Enceladus \citep[GSE;][]{Belokurov2018, Helmi2018}, and Sequoia\citep[e.g.,][]{Myeong2019}. Some of the more recent minor mergers between dwarf galaxies and the MW created magnificent observable streams; for example the Sagittarius (Sgr) streams \citep[e.g.,][]{Majewski2003, Law2010}. Therefore, stars formed in those dissolved and dissolving dwarf galaxies are important building blocks for our MW. 

While identifying dissolved dwarf galaxies is painstaking, existing dwarf galaxies are invaluable assets in that they allow us  to estimate not only the past but also the future evolution of our Galaxy, as they may be dragged into sufficient proximity by the Galactic potential in the future to merge with our MW. To better chemically tag dissolved dwarf galaxies, a full exploration of the chemistry of existing dwarf galaxies is required. One of the characteristic features of dwarf galaxies is their lower [$\alpha$/Fe] abundances compared to MW disc stars at a given [Fe/H] \citep{Tolstoy2009}. With the help of large spectroscopic surveys, studies of chemical evolution in dwarf galaxies are more efficient than ever before. Using high-resolution spectra from Apache Point Observatory Galactic Evolution Experiment \citep[APOGEE; ][]{Majewski2017},
\citet{Hasselquist2021} explored  the chemical abundances of multiple elements (C, N, O, Mg, Al, Si, Ca, Fe, Ni, and Ce) in the Large and Small Magellanic Clouds (LMC and SMC), Sgr dwarf, Fornax dwarf, and GSE. Chemical modeling of [$\alpha$/Fe]--[Fe/H] space reveals detailed star-formation epochs, and the possible existence of a second starburst.

The Sculptor (Scl) dwarf galaxy is a particular interesting example. Located at a distance of 86$\pm$3 kpc from the Sun \citep{Pietrzynski2008}, with M$_{V}=-10.82\pm$0.14 mag \citep{Munoz2018a}, Scl is considered as a textbook dwarf galaxy, with most stars older than 10 Gyr, and is therefore an ideal system with which to empirically investigate chemical evolution \citep[][hereafter H19]{Hill2019}.   As one of the nearby galaxies, Scl can be studied using a resolved color--magnitude diagram (CMD;  \citealt{Weisz2014}). Although much larger samples can be obtained than the spectroscopic ones, estimating the star-formation history from a resolved CMD may suffer from isochrone degeneracy at old ages ($>8$ Gyr) and low metallcity. On the other hand, observed chemical abundances combined with proper chemical evolution models (e.g., \citealt{2014MNRAS.441.2815V}, \citealt{2022ApJ...925...66D}) are powerful tools with which to reveal galaxy evolution. Using the medium-resolution spectra ($R\sim 6500$) from DEIMOS \citep{Faber2003} on the Keck II telescope, C, Mg, Si, Ca, Cr, Mn, Fe, Co, Ni, and Ba abundances of Scl stars were investigated in a series of works \citep{Kirby2010, Kirby2015,Kirby2018, Duggan2018, delosReyes2020}. The high-resolution spectra give more accurate chemical abundances, but obtaining these is time consuming. This is why  early high-resolution investigations in Scl have small sample sizes (of approximately five stars; e.g., \citealt{Shetrone2003,Geisler2005,Starkenburg2013,Jablonka2015,Simon2015}). The Scl sample size is increased to more than 100 in the DART survey \citep{Tolstoy2006}, which used ESO VLT/FLAMES to obtain high-resolution spectra ($R\sim 20,000$) of  red giant branch (RGB) stars. In the DART survey,
\citetalias{Hill2019} found a marked decrease in [$\alpha$/Fe] over their Scl sample, from the Galactic halo plateau value at low [Fe/H] and then, after a "knee", a decrease to subsolar [$\alpha$/Fe] at high [Fe/H]. The trend was consistent with products of core-collapse supernovae dominating at early times, followed by the onset of supernovae type Ia as early as $\sim$12 Gyr ago. 
By homogeneously analyzing chemical abundances of 380 stars in 13 dwarf galaxies (including Scl),  \citet[][hereafter R20]{Reichert2020} demonstrated that star formation history dominantly depends on galaxy mass. 

 Despite the wealth of studies on Scl, open key questions remain. For example, (1) it is still not clear as to whether Scl evolves differently in its radial direction. \citet{Bettinelli2019} found that the star formation history of Scl may be different across this dwarf galaxy: star formation in the innermost region was shown to last for longer ($\sim$ 1.5 Gyr) than that in the outermost region ($\sim$ 0.5 Gyr).  Given that star formation history can be traced by chemical abundances, we may be able to detect radial gradients of chemical abundances in Scl. Also, Al lines are relatively weak in optical for metal-poor stars, but it is important to discriminate (dissolved) dwarf galaxies from MW field stars \citep[e.g.,][]{Das2020}.    (2) It would also be helpful to identify a possible trend in Al abundances in Scl. 
In this work, we plan to tackle these questions by  investigating the chemical abundances derived from near-infrared (NIR) APOGEE spectra. 
The paper is organized as follows. In Section~\ref{sec:sample}, we describe our data and reduction processes. We also present a comparison between our derived elemental abundances and literature values. We compare Scl chemical abundances of O, Mg, Al, Si, Ca, Ti, Cr, Mn, Ni, Ce  with those of the MW and other dwarf galaxies in Section~\ref{sec:result}.  To reveal the chemical evolution of Scl behind our observations, we construct its galactic chemical evolution model and discuss its implications. We further discuss radial chemical gradients and N-rich field stars (Section  \ref{sec:discussion}). Our final conclusions are outlined in Section~\ref{sec:conclusion}.

\section{The Data}
\label{sec:sample}

\subsection{Observations and pipeline reduction}

Apache Point Observatory Galactic Evolution Experiment \citep[APOGEE,][]{Majewski2017} was one of the programs operating during the Sloan Digital Sky Survey III (SDSS-III, \citealt{Eisenstein2011}) and SDSS-IV \citep{Blanton2017}. Multi-object NIR fiber spectrographs observing from both the SDSS 2.5 m telescope at the Apache Point Observatory \citep{Gunn2006} and the 2.5 m du Pont telescope at the Las Campanas Observatory \citep{Bowen1973} deliver high-resolution ($R\sim$22,500) $H$-band spectra ($\lambda = 1.51 - 1.69$ $\mu$m). The APOGEE survey targeted a color-selected sample that predominantly consists of RGB stars across the Milky Way \citep{Zasowski2017, Santana2021, Beaton2021}.  APOGEE data-reduction software was applied to reduce multiple 3D raw data cubes into calibrated, well-sampled, combined 1D spectra \citep{Nidever2015}. The stellar parameters were derived using the FERRE code \citep{AllendePrieto2006}, which finds the best solution by comparing the observed spectra with libraries of theoretical spectra \citep{Zamora2015}.

\begin{figure*}
        \centering \includegraphics[width=1.0\linewidth,angle=0]{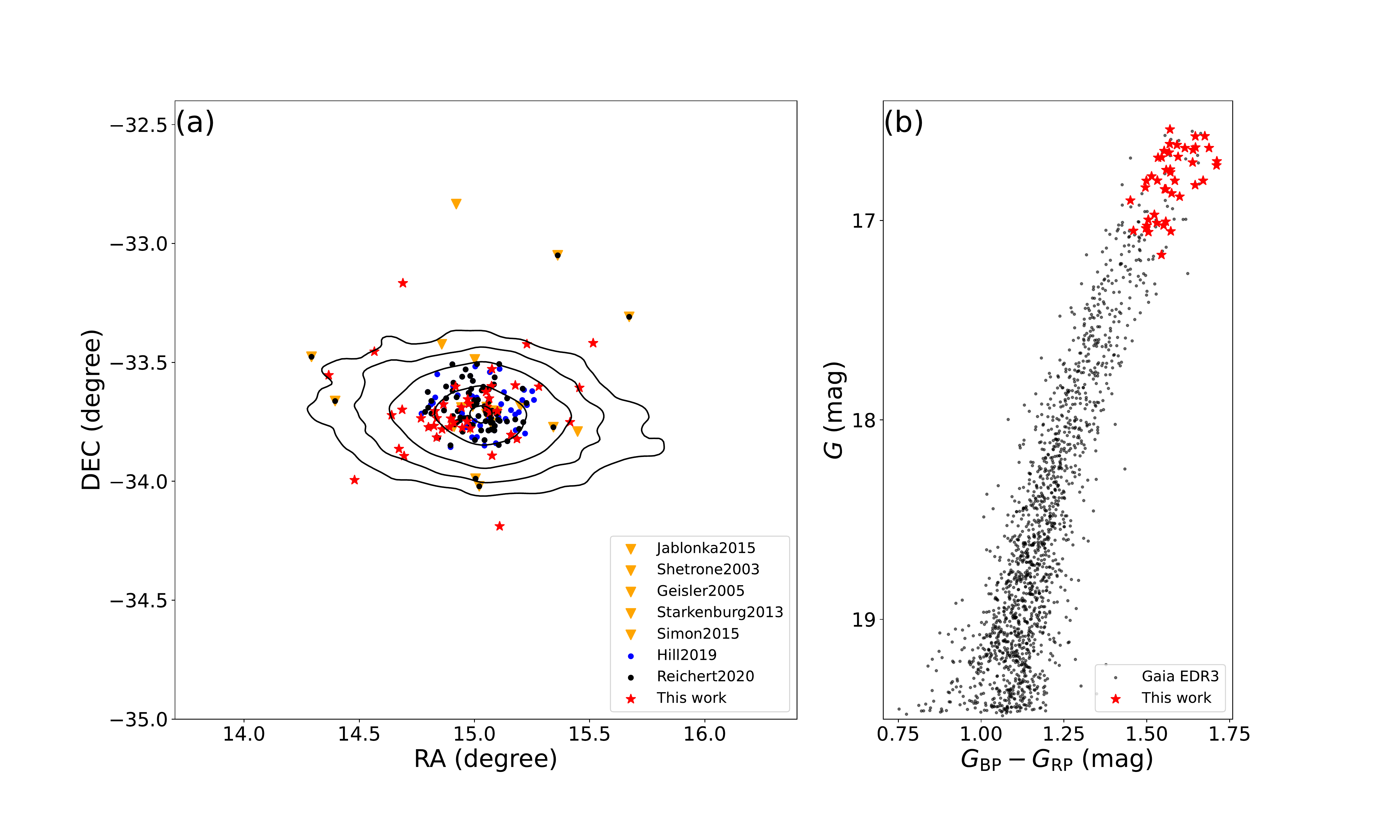}
        \caption{Target information. (a): Spatial distribution of observed Scl stars from the  literature. Scl stars from this work are colored red. Blue dots represent stars from \citetalias{Hill2019}. Black dots represent stars from \citetalias{Reichert2020}. Golden triangles show stars from \cite{Jablonka2015}, \cite{Shetrone2003}, \cite{Geisler2005}, \cite{Starkenburg2013}, and \cite{Simon2015}. An isodensity contour map represented by black lines is plotted based on Table 5 from \cite{Munoz2018b}. (b): Color--magnitude diagram of Scl based on $Gaia$ (gray dots).  }
        \label{fig:RADEC_CMD}
\end{figure*}

With 17  data releases of SDSS (DR 17; \citealt{Abdurrouf2022}) to date, the Scl dwarf galaxy has been observed multiple times, and the signal-to-noise ratios (S/Ns) of stacked spectra has reached a significant level. In this work, we cross-matched the Scl member candidates from \cite{HelmiGaia2018} with APOGEE DR17, and 43 stars are selected with S/N$>70$. APOGEE radial velocities (RVs) of these 43 stars are shown to be 112.4$\pm$8.9 km/s (Table~\ref{tab:stellar_parameters}), and all stars are within 3$\sigma$ limits. Therefore, our sample stars are highly likely Scl members based on their locations, proper motions, and RVs.
The locations  of our sample stars and their positions  on a color--magnitude diagram are plotted in Fig. \ref{fig:RADEC_CMD}. By comparing with previous studies, we find 15  stars in common with \citetalias{Hill2019}, and 12 in common with \citetalias{Reichert2020}.  These stars are useful for understanding the systematic errors of our measurements (see below).


\subsection{Stellar parameters and chemical abundances}
\label{subsec:sp}

 In this work, we calculated photometric $T_{\rm eff}$ from V band\footnote{\citet{Coleman2005}, kindly provided by Gary Da Costa.} and 2MASS photometry using the prescriptions described in \citet{RM05}. The final $T_{\rm eff}$ are the averages of $T_{\rm eff}$ derived from dereddened ($V-J$), ($V-H$), ($V-K$), where the extinction map is given by \citet{SFD1998}. Values for   $\log g$ were derived from $T_{\rm eff}$, stellar masses (0.8 solar mass is assumed for all our stars, given the age range of our sample), and bolometric luminosities, where the procedure outlined by \citet{Alonso1999} is used for bolometric correction. Our derived $T_{\rm eff}$ and  $\log g$ are shown in Table~\ref{tab:stellar_parameters}. 

\begin{table*}
        \caption{ Basic information on our sample stars}
        \label{tab:stellar_parameters}
        \begin{center}
                \scriptsize
\begin{tabular}{c c c c c c c c c c c}
\hline
\hline
\hline
        \multirow{2}{*}{\#} & \multirow{2}{*}{APOGEE\_ID} & RA & DEC &
        $T_{\rm eff}$ & \multirow{2}{*}{$\log g$}   & $RV$   & \multirow{2}{*}{S/N }  & Vmag  &Kmag    &   {\bf Common stars}  \\
                            &                     &          (deg)            &   ( deg) &
        (K)           &                   &${\rm \,(km\,s^{-1})}$ &   &(mag)       & (mag)    &      \\
\hline
\hline
 1  &  2M00572816-3333133  &  14.36736  &  -33.55371  &  4134.40  &  0.54  &  130.14  &  78.26  &  17.10  &  13.94  &    \\
 2  &  2M00575508-3359410  &  14.47952  &  -33.99473  &  4154.82  &  0.58  &  109.23  &  96.70  &  17.15  &  13.99  &    \\
 3  &  2M00581573-3327163  &  14.56557  &  -33.45455  &  4081.40  &  0.53  &  125.84  &  96.36  &  17.16  &  13.97  &    \\
 4  &  2M00583377-3343187  &  14.64074  &  -33.72189  &  4140.10  &  0.70  &  108.88  &  74.05  &  17.47  &  14.32  &    \\
 5  &  2M00584131-3351495  &  14.67213  &  -33.86376  &  4112.41  &  0.54  &  120.51  &  104.48  &  17.13  &  13.95  &    \\
 6  &  2M00584470-3341561  &  14.68628  &  -33.69893  &  4062.10  &  0.50  &  118.97  &  107.77  &  17.11  &  13.81  &    \\
 7  &  2M00584546-3310006  &  14.68944  &  -33.16685  &  4023.17  &  0.50  &  118.33  &  85.75  &  17.15  &  13.79  &    \\
 8  &  2M00584671-3353367  &  14.69464  &  -33.89354  &  4246.87  &  0.80  &  140.39  &  71.75  &  17.54  &  14.63  &    \\
 9  &  2M00590432-3344058  &  14.76802  &  -33.73496  &  4173.28  &  0.73  &  110.49  &  77.67  &  17.49  &  14.46  &    \\
 10  &  2M00591209-3346208  &  14.80041  &  -33.77246  &  4253.53  &  0.72  &  111.52  &  80.36  &  17.32  &  14.43  &  R20, H19  \\
 11  &  2M00591778-3346016  &  14.82411  &  -33.76712  &  4139.96  &  0.69  &  108.16  &  80.87  &  17.47  &  14.32  &    \\
 12  &  2M00591884-3342175  &  14.82854  &  -33.70488  &  4084.08  &  0.71  &  117.61  &  71.03  &  17.63  &  14.49  &  R20, H19  \\
 13  &  2M00592065-3348566  &  14.83607  &  -33.81574  &  4063.73  &  0.50  &  112.98  &  116.24  &  17.13  &  13.86  &  R20, H19  \\
 14  &  2M00592081-3344049  &  14.83672  &  -33.73470  &  4229.56  &  0.68  &  109.23  &  91.78  &  17.25  &  14.36  &  R20, H19  \\
 15  &  2M00592626-3346529  &  14.85945  &  -33.78136  &  4096.45  &  0.56  &  94.81  &  98.34  &  17.22  &  14.01  &    \\
 16  &  2M00592768-3340356  &  14.86537  &  -33.67656  &  4146.20  &  0.58  &  119.66  &  94.59  &  17.17  &  14.03  &  H19  \\
 17  &  2M00593400-3346210  &  14.89168  &  -33.77252  &  4139.87  &  0.71  &  107.25  &  74.42  &  17.51  &  14.34  &    \\
 18  &  2M00593538-3344094  &  14.89743  &  -33.73597  &  3841.07  &  0.34  &  111.25  &  115.04  &  17.31  &  13.62  &    \\
 19  &  2M00593841-3345107  &  14.91005  &  -33.75299  &  4091.02  &  0.68  &  116.77  &  77.19  &  17.52  &  14.24  &    \\
 20  &  2M00594032-3336064  &  14.91802  &  -33.60178  &  4095.91  &  0.68  &  112.45  &  75.22  &  17.53  &  14.49  &    \\
 21  &  2M00594641-3341233  &  14.94341  &  -33.68982  &  3908.62  &  0.39  &  107.26  &  110.67  &  17.25  &  13.75  &  H19, Geisler2005  \\
 22  &  2M00594727-3346305  &  14.94696  &  -33.77516  &  4029.68  &  0.63  &  106.54  &  80.78  &  17.54  &  14.25  &    \\
 23  &  2M00595228-3344546  &  14.96784  &  -33.74850  &  4105.40  &  0.61  &  117.41  &  94.57  &  17.32  &  14.07  &  H19  \\
 24  &  2M00595302-3339189  &  14.97092  &  -33.65527  &  3964.64  &  0.51  &  111.28  &  89.88  &  17.40  &  14.01  &    \\
 25  &  2M00595422-3340270  &  14.97593  &  -33.67418  &  4068.80  &  0.65  &  99.17  &  77.54  &  17.51  &  14.36  &  R20, H19  \\
 26  &  2M00595568-3346400  &  14.98202  &  -33.77778  &  4182.28  &  0.65  &  121.84  &  83.97  &  17.28  &  14.25  &  R20, H19, Geisler2005  \\
 27  &  2M01001215-3337260  &  15.05064  &  -33.62389  &  4107.19  &  0.50  &  113.06  &  109.85  &  17.01  &  13.81  &    \\
 28  &  2M01001278-3341157  &  15.05328  &  -33.68772  &  4197.29  &  0.66  &  97.31  &  78.73  &  17.26  &  14.30  &  R20, H19, Shetrone2003  \\
 29  &  2M01001537-3339060  &  15.06407  &  -33.65169  &  4071.09  &  0.58  &  112.78  &  88.81  &  17.33  &  14.20  &  R20, H19  \\
 30  &  2M01001630-3342371  &  15.06795  &  -33.71033  &  3959.43  &  0.38  &  103.56  &  110.03  &  17.07  &  13.70  &  Geisler2005  \\
 31  &  2M01001777-3335597  &  15.07408  &  -33.59993  &  3968.14  &  0.49  &  120.20  &  93.51  &  17.34  &  13.95  &  R20, H19  \\
 32  &  2M01001821-3331405  &  15.07590  &  -33.52794  &  3978.96  &  0.54  &  105.27  &  95.09  &  17.41  &  13.93  &    \\
 33  &  2M01001833-3353314  &  15.07641  &  -33.89208  &  4112.65  &  0.62  &  128.62  &  81.15  &  17.34  &  14.14  &    \\
 34  &  2M01002384-3342173  &  15.09937  &  -33.70483  &  4073.31  &  0.58  &  122.14  &  77.80  &  17.33  &  14.05  &  R20, H19, Geisler2005  \\
 35  &  2M01002624-3411206  &  15.10937  &  -34.18906  &  4076.53  &  0.60  &  122.50  &  87.82  &  17.36  &  14.09  &    \\
 36  &  2M01003813-3348168  &  15.15889  &  -33.80468  &  4071.18  &  0.55  &  107.07  &  89.94  &  17.25  &  13.92  &  R20, H19  \\
 37  &  2M01004256-3335475  &  15.17736  &  -33.59654  &  4139.44  &  0.55  &  113.03  &  96.69  &  17.10  &  13.98  &    \\
 38  &  2M01004427-3349188  &  15.18449  &  -33.82190  &  4053.48  &  0.64  &  112.67  &  71.51  &  17.52  &  14.32  &  R20, H19  \\
 39  &  2M01005465-3325231  &  15.22775  &  -33.42311  &  4036.26  &  0.56  &  104.17  &  86.07  &  17.36  &  14.06  &    \\
 40  &  2M01010709-3336065  &  15.27957  &  -33.60181  &  4049.61  &  0.52  &  107.91  &  90.37  &  17.23  &  14.01  &    \\
 41  &  2M01013961-3345043  &  15.41507  &  -33.75120  &  3993.39  &  0.45  &  103.55  &  95.28  &  17.16  &  13.78  &    \\
 42  &  2M01014939-3336241  &  15.45583  &  -33.60670  &  4003.54  &  0.50  &  113.29  &  93.68  &  17.17  &  13.87  &    \\
 43  &  2M01020368-3325079  &  15.51535  &  -33.41888  &  4091.18  &  0.55  &  99.55  &  82.42  &  17.21  &  13.94  &    \\
\hline
\hline
\hline
\end{tabular}
        \end{center}
        \raggedright{\hspace{0cm} Note: R20-\cite{Reichert2020}, H19-\cite{Hill2019}, Geisler2005-\cite{Geisler2005}, Shetrone2003-\cite{Shetrone2003}}.\\
\end{table*}

Local thermal equilibrium (LTE) chemical abundances (including metallicity from Fe I lines) were analyzed using Brussels Automatic Stellar Parameter (BACCHUS) code \citep{Masseron2016}, where careful line selection and detailed line inspection are performed. The BACCHUS code is developed on top of the radiative transfer code Turbospectrum \citep{Alvarez1998, Plez2012}, and model atmosphere is interpolated from  MARCS  model grids \citep{Gustafsson2008}.
To derive chemical abundance for a given absorption line, a sigma-clipping is first applied on the selected continuum points around the targeted line, and then a linear fit is used for the remaining points as the continuum. Therefore, the code can detect significantly poor fits, such as a sudden drop in the spectrum due to bad pixels in the detectors. Observed spectra and model spectra are compared with four different methods to determine abundances: 1) $\chi^2$ minimization  of global goodness-of-fit; 2) core line intensity comparison; 3) equivalent width comparison, and 4) spectral synthesis (see Appendix Fig. in \citealt{Yu2021}). Furthermore, the code gives each of them a flag to indicate the estimation quality, which is used to reject or accept the line, keeping the best-fit abundance.  To increase the credibility of our measurements, we only accept a line when flags indicate the fittings are good in all four methods. If no line meets the above criteria, the abundance is not estimated.

Across the wavelength range of APOGEE spectra, molecular lines of CO, CN, and OH could be significant features affecting the line-by-line abundance analysis. Therefore, we estimated the C, N, O, and Fe abundances self-consistently: we first derived $^{16}$O abundances from $^{16}$OH lines; we then derived $^{12}$C from $^{12}$C$^{16}$O lines, and $^{14}$N from $^{12}$C$^{14}$N; and subsequently, iron abundances were derived from Fe I lines. This process was iterated to minimize the dependence of the molecular lines and Fe I lines \citep{Smith2013}. 
After that, atomic lines of Mg, Al, Si, Ca, Ti, Cr, Mn, Ni, and Ce were analyzed line-by-line.   Mg, Al, and Si show strong atomic lines in the observed wavelength, and therefore the abundances of these elements are mostly complete across our sample stars. On the other hand, a large number of Cr, Mn, Ni, and Ce atomic lines and $^{12}$C$^{16}$O and $^{12}$C$^{14}$N molecular lines are too weak to give meaningful results; these were therefore rejected visually to avoid false detection. 
 The derived chemical abundances of our targets are given in Table~\ref{tab:abundances}. We assumed solar abundances of \citet{Asplund2005} in this work.
Other chemical elements mostly show weak atomic features in the observed wavelength (e.g., Na), which is partially because of the metal-poor nature of our stars ([Fe/H$]<-1$), and these are not discussed here. 

\begin{table*}
        \caption{Chemical abundances}
        \label{tab:abundances}
        \begin{center}
                \scriptsize
\begin{tabular}{c c c c c c c c c c c c c c}
\hline
\hline
\hline
 \#  &   [Fe/H]      &   [C/Fe]  &   [N/Fe]  &   [O/Fe]  &   [Mg/Fe]  &   [Al/Fe]  &   [Si/Fe]  &   [Ca/Fe]  &   [Ti/Fe]  &   [Cr/Fe]  &   [Mn/Fe]  &   [Ni/Fe]  &   [Ce/Fe]  \\ 
\hline
\hline
1&    -1.75  &  ...  &  ...  &     0.34  &    -0.01  &  ...  &     0.07  &  ...  &  ...  &  ...  &  ...  &  ...  &  ...  \\ 
2&    -2.16  &  ...  &  ...  &     0.25  &     0.30  &    -0.55  &     0.35  &     0.61  &  ...  &  ...  &  ...  &  ...  &  ...  \\ 
3&    -1.91  &  ...  &  ...  &     0.31  &     0.05  &    -0.38  &     0.22  &     0.48  &  ...  &  ...  &  ...  &  ...  &  ...  \\ 
4&    -1.92  &  ...  &  ...  &     0.67  &     0.12  &    -0.55  &     0.22  &     0.66  &  ...  &  ...  &  ...  &  ...  &  ...  \\ 
5&    -2.01  &  ...  &  ...  &     0.23  &    -0.15  &    -0.55  &     0.22  &     0.47  &  ...  &  ...  &  ...  &  ...  &  ...  \\ 
6&    -2.13  &  ...  &  ...  &     0.21  &     0.14  &    -0.52  &     0.19  &  ...  &  ...  &  ...  &  ...  &  ...  &  ...  \\ 
7&    -1.64  &     0.40  &     0.86  &     0.34  &    -0.02  &    -0.69  &     0.08  &  ...  &  ...  &  ...  &  ...  &  ...  &     0.64  \\ 
8&    -1.81  &  ...  &  ...  &     0.31  &     0.08  &    -0.24  &     0.13  &  ...  &  ...  &  ...  &  ...  &  ...  &  ...  \\ 
9&    -1.60  &  ...  &  ...  &     0.23  &  ...  &    -0.67  &     0.04  &    -0.01  &    -0.13  &  ...  &  ...  &  ...  &  ...  \\ 
10&    -1.88  &     0.57  &     1.42  &     0.75  &  ...  &    -0.46  &     0.13  &  ...  &  ...  &  ...  &  ...  &  ...  &  ...  \\ 
11&    -1.59  &  ...  &  ...  &     0.19  &    -0.31  &    -0.63  &     0.05  &     0.26  &  ...  &  ...  &  ...  &  ...  &  ...  \\ 
12&    -1.22  &  ...  &  ...  &     0.05  &    -0.57  &  ...  &     0.03  &    -0.12  &    -0.47  &  ...  &  ...  &  ...  &  ...  \\ 
13&    -1.76  &  ...  &  ...  &     0.32  &     0.05  &    -0.44  &     0.22  &     0.34  &     0.16  &  ...  &  ...  &  ...  &  ...  \\ 
14&    -1.91  &  ...  &  ...  &     0.35  &    -0.28  &    -0.67  &     0.09  &  ...  &  ...  &  ...  &  ...  &  ...  &  ...  \\ 
15&    -1.94  &    -0.17  &  ...  &     0.25  &    -0.03  &    -0.63  &     0.11  &     0.05  &  ...  &  ...  &  ...  &  ...  &  ...  \\ 
16&    -2.09  &  ...  &  ...  &     0.23  &     0.08  &    -0.58  &     0.20  &     0.12  &  ...  &  ...  &  ...  &  ...  &  ...  \\ 
17&    -1.61  &  ...  &  ...  &     0.34  &    -0.02  &  ...  &     0.18  &     0.43  &  ...  &  ...  &  ...  &  ...  &  ...  \\ 
18&    -1.05  &    -1.10  &  ...  &    -0.21  &    -0.44  &    -1.10  &    -0.03  &    -0.31  &    -0.70  &    -0.22  &    -0.47  &  ...  &    -0.33  \\ 
19&    -1.59  &  ...  &  ...  &     0.08  &    -0.07  &    -0.41  &     0.18  &     0.17  &  ...  &  ...  &  ...  &     0.16  &  ...  \\ 
20&    -1.57  &  ...  &  ...  &     0.21  &    -0.03  &    -0.68  &     0.09  &     0.24  &     0.40  &  ...  &  ...  &     0.18  &  ...  \\ 
21&    -1.05  &    -0.58  &  ...  &    -0.00  &    -0.36  &    -0.94  &    -0.03  &    -0.07  &    -0.62  &     0.03  &    -0.29  &    -0.38  &    -0.25  \\ 
22&    -1.15  &    -0.63  &  ...  &    -0.02  &    -0.31  &    -1.06  &    -0.03  &    -0.15  &     0.16  &  ...  &    -0.41  &  ...  &    -0.24  \\ 
23&    -2.05  &  ...  &  ...  &     0.29  &     0.04  &    -0.55  &     0.19  &    -0.09  &  ...  &  ...  &  ...  &  ...  &  ...  \\ 
24&    -1.14  &    -1.16  &  ...  &    -0.10  &    -0.48  &    -0.88  &    -0.00  &    -0.08  &  ...  &     0.12  &    -0.21  &  ...  &  ...  \\ 
25&    -1.30  &    -0.52  &  ...  &     0.10  &  ...  &    -0.34  &    -0.04  &     0.07  &     0.29  &  ...  &    -0.11  &  ...  &  ...  \\ 
26&    -2.14  &  ...  &  ...  &     0.32  &     0.02  &    -0.60  &     0.13  &  ...  &  ...  &  ...  &  ...  &  ...  &  ...  \\ 
27&    -1.65  &  ...  &  ...  &     0.38  &    -0.06  &    -0.49  &     0.16  &     0.15  &     0.14  &  ...  &  ...  &  ...  &     0.19  \\ 
28&    -1.73  &    -0.12  &  ...  &     0.30  &    -0.06  &    -0.63  &     0.05  &  ...  &  ...  &  ...  &  ...  &  ...  &  ...  \\ 
29&    -1.42  &  ...  &  ...  &     0.05  &    -0.10  &  ...  &     0.16  &  ...  &  ...  &  ...  &  ...  &    -0.08  &  ...  \\ 
30&    -1.43  &     0.40  &     1.61  &     0.37  &  ...  &    -0.69  &     0.17  &  ...  &    -0.28  &  ...  &  ...  &  ...  &     2.23  \\ 
31&    -1.13  &    -0.76  &  ...  &    -0.05  &  ...  &    -1.18  &     0.02  &    -0.04  &    -0.33  &    -0.03  &    -0.32  &  ...  &    -0.23  \\ 
32&    -1.63  &  ...  &  ...  &     0.32  &     0.11  &    -0.50  &     0.31  &  ...  &  ...  &  ...  &  ...  &     0.26  &  ...  \\ 
33&    -1.72  &  ...  &  ...  &     0.33  &    -0.10  &    -0.53  &     0.15  &  ...  &    -0.08  &  ...  &  ...  &  ...  &  ...  \\ 
34&    -1.71  &  ...  &  ...  &     0.14  &    -0.11  &    -0.42  &     0.30  &  ...  &    -0.22  &  ...  &  ...  &  ...  &  ...  \\ 
35&    -1.86  &  ...  &  ...  &     0.19  &     0.11  &    -0.26  &     0.16  &    -0.04  &  ...  &  ...  &  ...  &  ...  &  ...  \\ 
36&    -1.76  &  ...  &  ...  &     0.20  &     0.03  &    -0.33  &     0.16  &    -0.03  &  ...  &  ...  &  ...  &  ...  &  ...  \\ 
37&    -1.81  &  ...  &  ...  &     0.33  &    -0.06  &  ...  &     0.06  &  ...  &  ...  &  ...  &  ...  &  ...  &  ...  \\ 
38&    -1.32  &  ...  &  ...  &    -0.01  &  ...  &    -0.50  &     0.11  &     0.26  &     0.35  &  ...  &  ...  &    -0.46  &  ...  \\ 
39&    -1.55  &  ...  &  ...  &     0.21  &    -0.06  &    -0.53  &     0.19  &     0.09  &  ...  &  ...  &  ...  &  ...  &  ...  \\ 
40&    -1.57  &    -0.10  &  ...  &     0.06  &  ...  &  ...  &    -0.08  &    -0.03  &  ...  &  ...  &  ...  &  ...  &  ...  \\ 
41&    -1.56  &  ...  &  ...  &     0.17  &    -0.04  &    -0.50  &     0.15  &     0.10  &  ...  &  ...  &  ...  &     0.10  &     0.00  \\ 
42&    -1.33  &    -0.80  &  ...  &     0.03  &  ...  &    -1.00  &     0.03  &     0.05  &    -0.53  &    -0.44  &  ...  &  ...  &    -0.08  \\ 
43&    -2.05  &  ...  &  ...  &     0.16  &     0.04  &    -0.49  &     0.24  &     0.45  &     0.68  &  ...  &  ...  &  ...  &  ...  \\ 

\hline
\hline
\end{tabular}
        \end{center}
\end{table*}

We analyzed the internal errors by propagating typical errors in $T_{\rm eff}$, $\log g$, and ${\rm [Fe/H]}$. The typical errors are set as $\Delta T_{\rm eff} = 100\,{\rm K}$, $\Delta \log g = 0.25\,{\rm dex}$,  and $\Delta {\rm [Fe/H]} = 0.1\,{\rm dex}$. The total estimated error is therefore calculated as: $\sigma_{\rm tot} = \left( (\sigma_{\Delta T_{\rm eff}})^2 + (\sigma_{\Delta \log g})^2 + (\sigma_{\Delta {\rm [Fe/H]}})^2 \right)^{1/2}$. The abundance errors of a typical star (\#21) are shown in Table~\ref{tab:abu_err} as an example.

\begin{table*}
    \caption{Errors of chemical abundances propagated from atmospheric parameters for star \#21}
        \label{tab:abu_err}
        \begin{center}
                \scriptsize
\begin{tabular}{c c c c c}
\hline
\hline
\hline
        Element  & $\Delta T_{\rm eff}=100 {\rm \,(K)}$ & $\Delta \log (g) = 0.25$\,(dex) & $\Delta {\rm [Fe/H]} = 0.1$\,(dex) & $\sigma_{\rm tot}$ \\
\hline
\hline
{\rm [Fe/H]} &  0.022  &  0.171  & 0.012  & 0.173  \\ 
{\rm [C/Fe]} &  0.272  &  0.515  & 0.250  & 0.634  \\ 
{\rm [N/Fe]} &  ...  &  ...  & ...  & ...  \\ 
{\rm [O/Fe]} &  0.237  &  0.080  & 0.002  & 0.250  \\ 
{\rm [Mg/Fe]} &  ...  &  0.185  & ...  & ...  \\ 
{\rm [Al/Fe]} &  0.080  &  0.046  & 0.010  & 0.093  \\ 
{\rm [Si/Fe]} &  0.002  &  0.011  & 0.012  & 0.016  \\ 
{\rm [Ca/Fe]} &  0.049  &  0.177  & 0.015  & 0.185  \\ 
{\rm [Ti/Fe]} &  1.010  &  0.053  & 0.883  & 1.343  \\ 
{\rm [Cr/Fe]} &  0.111  &  0.201  & 0.018  & 0.230  \\ 
{\rm [Mn/Fe]} &  0.007  &  0.112  & 0.020  & 0.114  \\ 
{\rm [Ni/Fe]} &  0.002  &  0.018  & 0.043  & 0.047  \\ 
{\rm [Ce/Fe]} &  0.040  &  0.027  & 0.024  & 0.053  \\ 

\hline
\hline
\hline
\end{tabular}

        \end{center}
\end{table*}

\subsection{Comparison with literature}
\label{subsec:cp}

\begin{figure*}
        \centering \includegraphics[width=0.9\linewidth,angle=0]{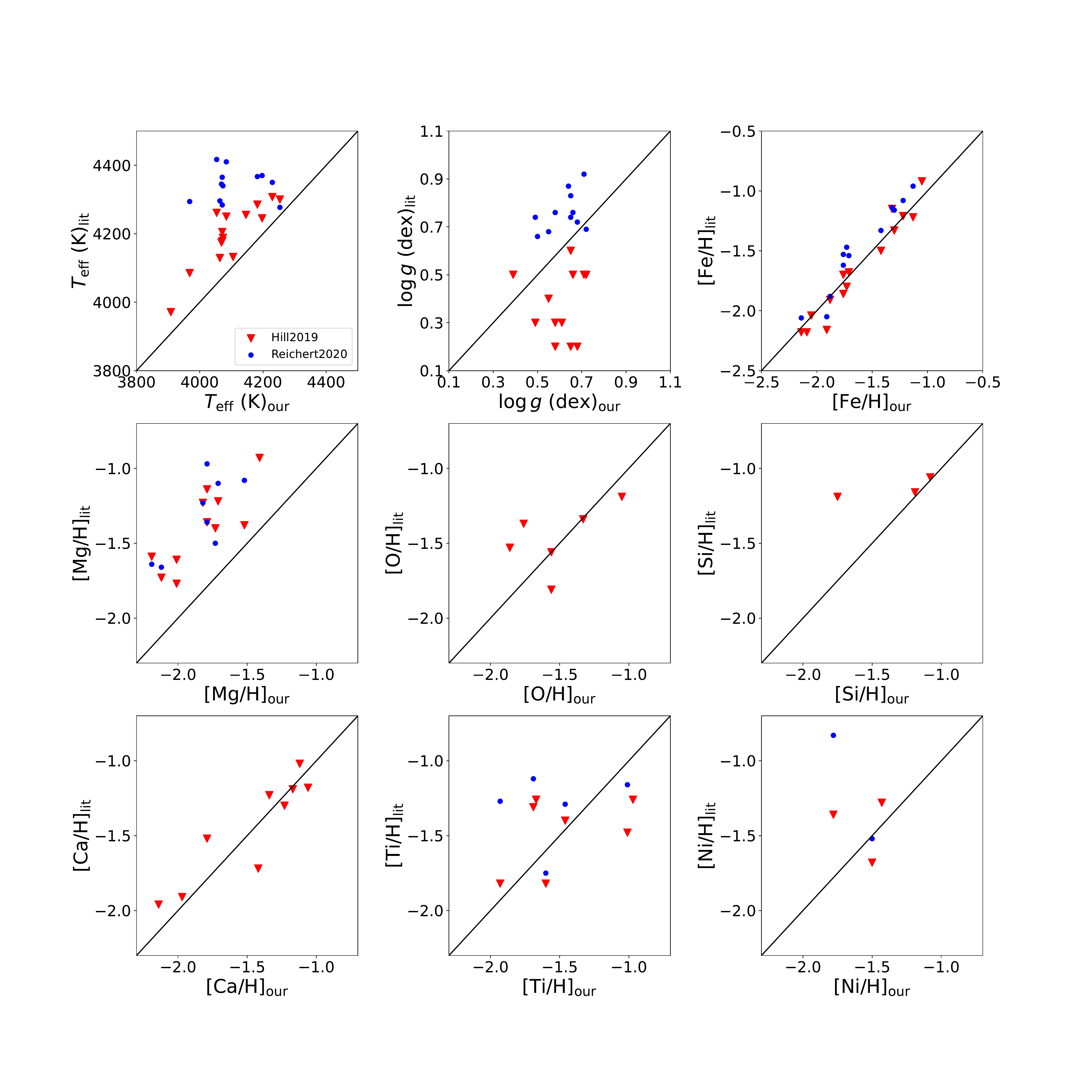}
        \caption{Comparison of stellar atmospheric parameters and elemental abundances of this work with those from the literature. Our results and those from the literature are denoted ``our'' and ``lit'', respectively. Red triangles represent comparisons between our work and \citetalias{Hill2019} and blue circles represent comparisons between our work and \citetalias{Reichert2020}. The black solid line is the 1:1 relation. 
        }
        \label{fig:compare_abu}
\end{figure*}

As our study is the first to investigate Scl chemical abundances in NIR, it is important to investigate the possible systematic errors associated with our approach by comparing our results with those obtained from optical spectra. 
  The upper panels of  Fig.
\ref{fig:compare_abu}  show a comparison of the stellar parameters $T_{\rm eff}$, $\log g$, and [Fe/H] from our work with those derived from optical spectra in the literature. Given that \citetalias{Hill2019} adopted photometric $T_{\rm eff}$, and \citetalias{Reichert2020} adopted spectroscopic $T_{\rm eff}$, our $T_{\rm eff}$ agree better with those of \citetalias{Hill2019}, but there is still an offset ($\sim$100 K). In \citetalias{Hill2019}, $T_{\rm eff}$ were derived by averaging over $T_{\rm eff}$ from ($V-I$), ($V-J$), ($V-K$), where $T_{\rm eff}$ determined from ($V-I$) are on average 200 K hotter than those from ($V-J$) and ($V-K$). As most of our sample stars are outside the application range of the $(V-I) - T_{\rm eff}$ relation, we chose not to use $T_{\rm eff}$ determined from ($V-I$), and therefore our adopted $T_{\rm eff}$ are cooler than those of \citetalias{Hill2019}. Interestingly, our adopted $T_{\rm eff}$ show no significant systematic offset compared to the uncalibrated ASPCAP $T_{\rm eff}$, with a small difference scatter ($\sim$ 25 K).
 The $\log g$ from our work are generally smaller than those of \citetalias{Reichert2020}, which is related to the cooler $T_{\rm eff}$ that we adopted. However, our $\log g$ are larger than those derived in \citetalias{Hill2019}, which may be a result of adopting spectroscopic $\log g$ in \citetalias{Hill2019}. 
\,\,\,[Fe/H] derived in this work is consistent with that of \citetalias{Hill2019}, while being  slightly more metal poor than that of \citetalias{Reichert2020}, which is related to the cooler $T_{\rm eff}$ and smaller $\log g$ in this work.

We compare the element abundances that we find here with those from other optical studies in Fig. \ref{fig:compare_abu}. For [O/H], [Si/H], [Ca/H], and [Ni/H], most of the data points are located on both sides of the 1:1 line, although with large scatter or low number statistics. For example, O and Si are measured in all our sample stars, but they were only measured in a small portion of the star sample of \citetalias{Hill2019}, and none were measured in \citetalias{Reichert2020}. This is due to the stronger OH molecular lines and Si atomic lines in the NIR. However, the opposite is true for Ni.  The large scatter in Ti is caused by the large derivation uncertainties (see Table \ref{tab:abu_err}) due to the relative weakness of the atomic lines in APOGEE spectra.
For Mg, we find an under-abundance of $\sim$0.4 dex compared to the literature.
The nonLTE (NLTE) corrections for optical Mg lines are relatively small in the metallicity range  that we are interested in ($-2<[$Fe/H$]<-1$; see Fig.6 of \citetalias{Reichert2020}). Meanwhile,
according to NLTE investigations of Mg lines in the APOGEE wavelengths \citep[e.g.,][]{Zhang2017,Osorio2020,Lind2022}, the NLTE correction for Mg could be negligible when the spectral resolution is sufficiently high ($R \sim 100,000$). However, lower resolution spectra may not resolve the cores of the lines adequately, resulting in a substantial NLTE correction ($0.2-0.4$ dex) for the prominent Mg triplet ($\lambda \sim 15741, 15749, 15766$ \AA) in the APOGEE spectra \citep{Osorio2020,Masseron2021}. The Mg abundances in this work are in good agreement with those from the APOGEE DR17 pipeline. The aforementioned Mg under-abundance should be common for metal-poor giant stars when comparing APOGEE-derived abundances with optical ones.
 The comparison here reveals a caveat that must be taken into account when interpreting our results.
 

\section{Results}
\label{sec:result}

\subsection{\texorpdfstring{$\alpha$}{alpha}-elements}
\label{sec:alpha}

The $\alpha$-elements are mainly generated in massive stars through type II supernovae (SNe II), and consequently their recycle timescale in the interstellar medium is much smaller than that of iron, which is mainly produced in type Ia SNe (SNe Ia). As a result, the $[\alpha/{\rm Fe}]$ abundance is initially enhanced, and starts to decline with [Fe/H] after the SNe Ia delay time \citep{Matteucci1986}. The declining point, or the ``knee'' in $[\alpha/{\rm Fe}]$ versus ${\rm [Fe/H]}$ trend, is related to the star formation rate and therefore galaxy mass: the less massive the galaxy, the more metal-poor the $[\alpha/{\rm Fe}]$ turnover. 

\begin{figure*}
        \centering \includegraphics[width=0.7\linewidth,angle=0]{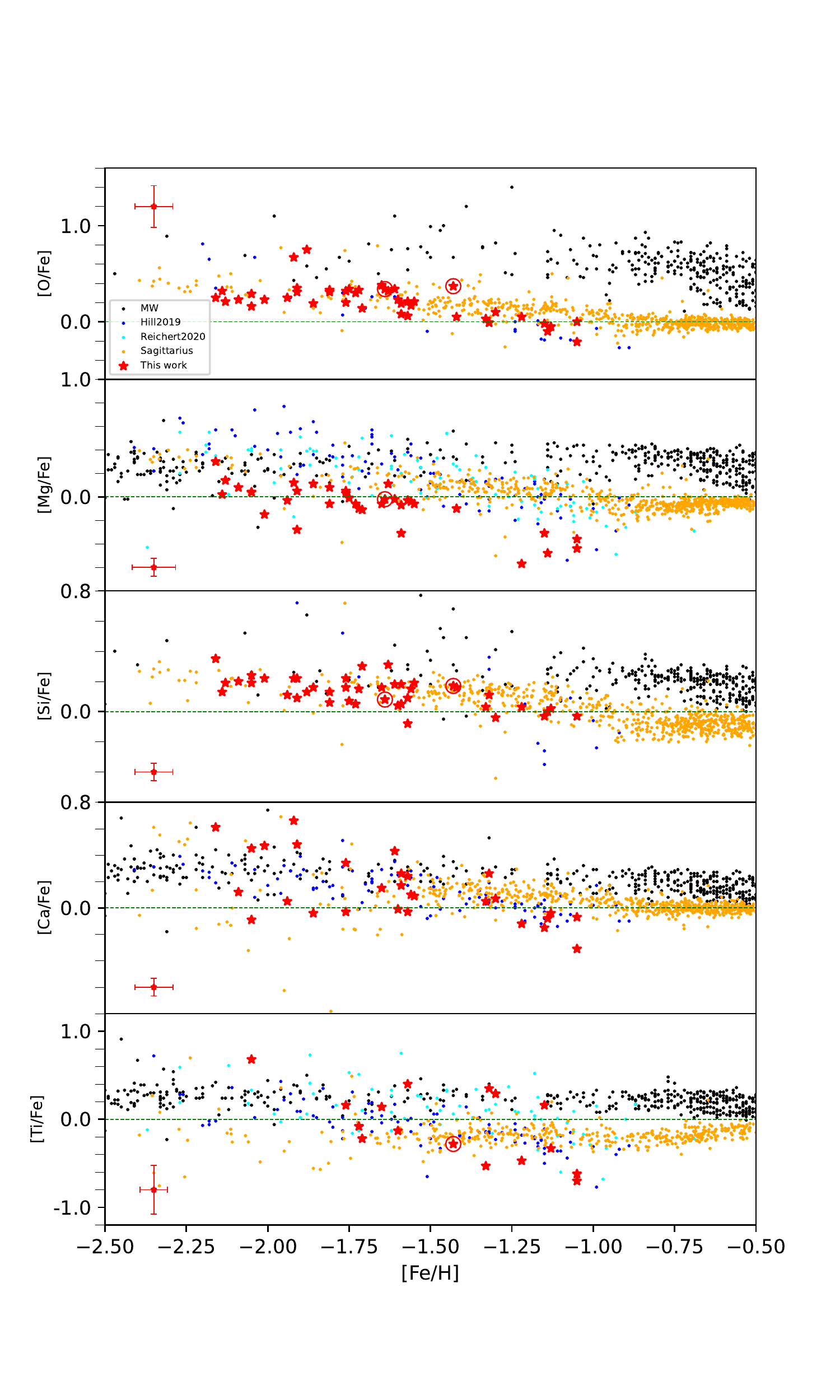}
        \caption{[$\alpha$/Fe] vs. [Fe/H]. Scl stars from this work are labeled as red stars. The error bar in each panel indicates the median uncertainty of available measurements. Possible AGB stars are outlined with large circles. Black dots correspond to MW stars from the halo (92 out of 168 stars from \citealt{Fulbright2000}, 35 stars from \citealt{Cayrel2004}, 234 out of 253 stars from \citealt{Barklem2005}, 131 out of 199 stars from \citealt{Yong2013}, 287 out of 313 stars from \citealt{Roederer2014}),  and MW stars from the disc (174 out of 181 stars from \citealt{Reddy2003}, 153 out of 176 stars from \citealt{Reddy2006}, 679 out of 714 stars from \citealt{Bensby2014}). Blue and cyan dots represent Scl stars from \citetalias{Hill2019} and \citetalias{Reichert2020}, respectively.  Orange dots represent Sgr stars from \cite{Hasselquist2021}. }
        \label{fig:alpha_FeH}
\end{figure*}

The [$\alpha$/Fe]-[Fe/H] relations are shown in Fig.~\ref{fig:alpha_FeH}. In the background, we also include MW halo, disc, and  dwarf galaxy stars\footnote{ To maintain the legibility of each figure, but still grasp the essence of abundance changes with respect to galaxy mass, we only select the two most representative and well-studied examples: MW and Sgr. For chemical abundances of other nearby dwarf galaxies, please refer to \citetalias{Reichert2020} or \citet{Hasselquist2021}.}. 
[$\alpha$/Fe] of our sample stars all show declining trends as a function of metallicity.  We note that (1) this is the first time that high-resolution O and Si abundances have been investigated with such a large sample size in Scl, which is thanks to their strong feature lines in the NIR;
 and that (2) Ti lines in the observed spectra are relatively weak, especially for more metal-poor stars. As a result, our Ti abundances show larger scatter than those of \citetalias{Hill2019} and \citetalias{Reichert2020}. 

 In terms of galaxy mass, the MW is much larger than Sgr and Scl, and we confirm that MW stars show the largest [$\alpha$/Fe] element abundances among them.
Comparing [$\alpha$/Fe] of our study and those of Sgr, where similar NIR lines were used to derive abundances, we find that Scl stars of this work show systematically lower [$\alpha$/Fe] than Sgr stars, particularly at [Fe/H]>-1.5. This is more robust for O, Mg, and Si, but is less evident for Ca and Ti. Our observations qualitatively agree with the theory that less massive galaxies generate less $\alpha$ elements. 


\subsection{Al}

Al is mainly synthesized by C and Ne burning in massive stars \citep{Arnett1971, Woosley1995}, but can also be modified by the Mg-Al cycle activated at higher core temperatures during the asymptotic giant branch (AGB) H-burning phase \citep{Arnould1999}. However, the nonexistence of a Mg-Al anti-correlation and the small scatter of [Mg/Fe] at a given metallicity suggest that the Mg-Al nucleosynthesis cycle does not significantly change their abundances; that is, the Al abundances of our sample reflect their primordial chemistry.

\begin{figure*}
        \centering \includegraphics[width=0.7\linewidth,angle=0]{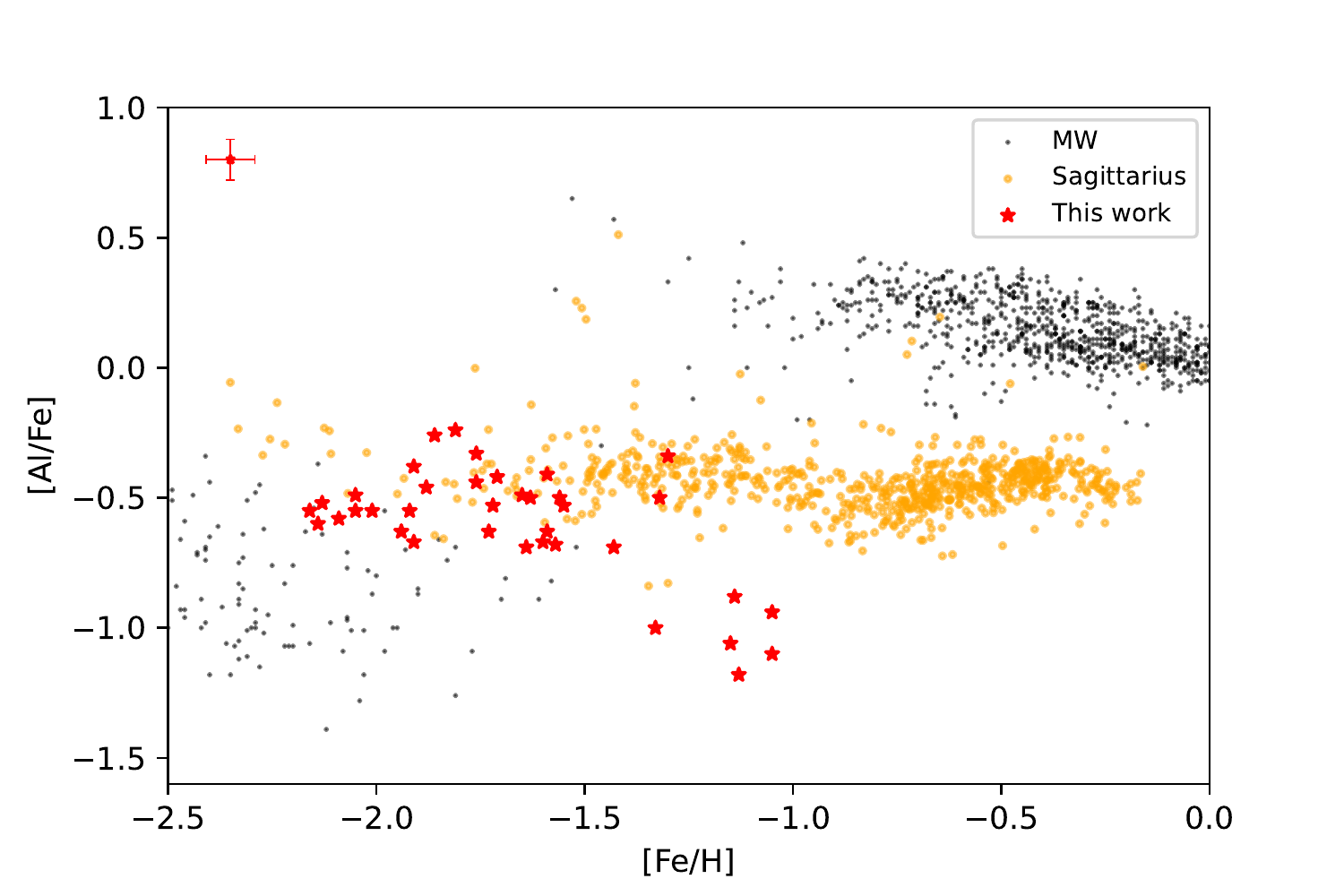}
        \centering \includegraphics[width=0.7\linewidth,angle=0]{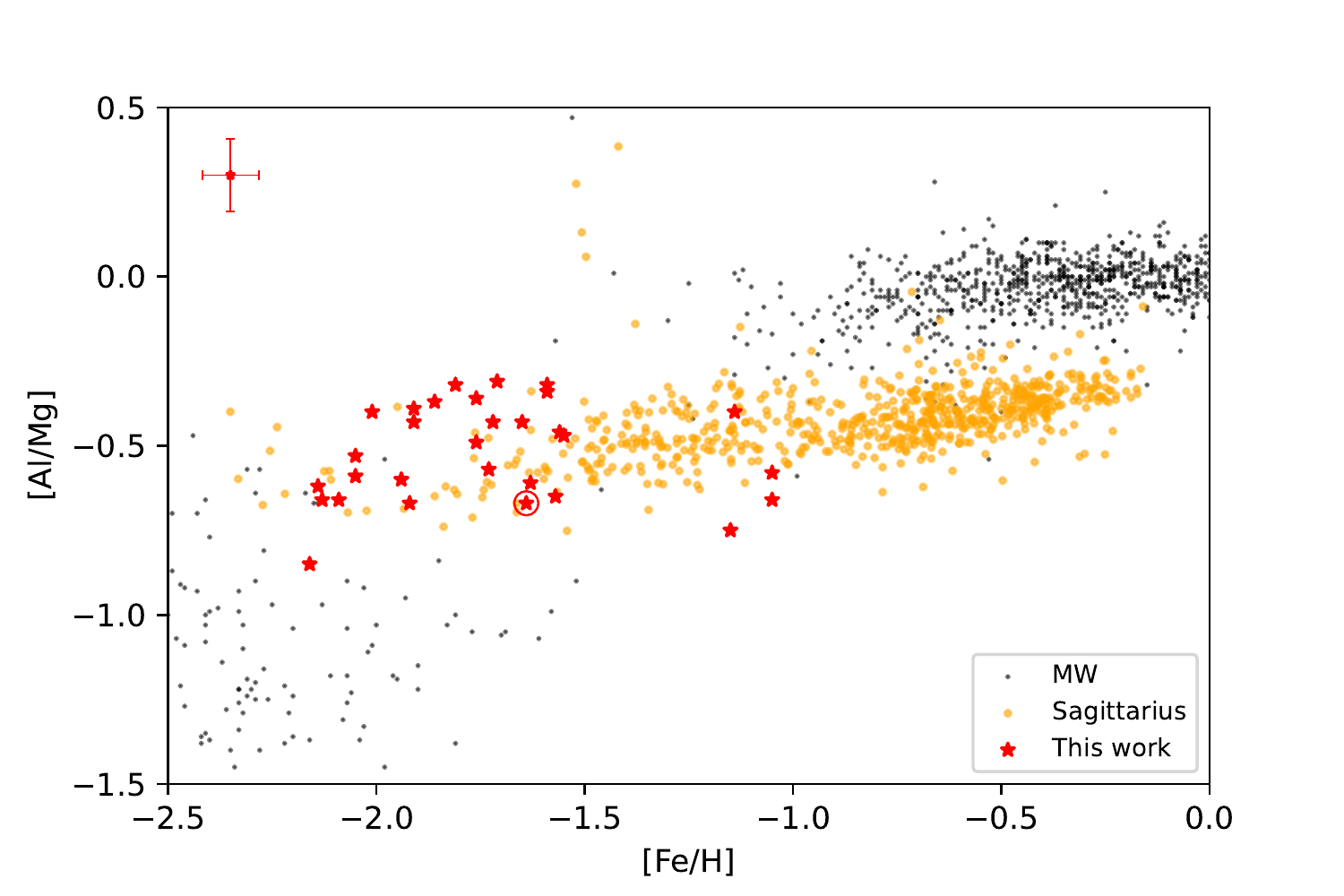}
        \caption{[Al/Fe] ([Al/Mg]) vs. [Fe/H] relations. Symbols are the same as in Fig.~\ref{fig:alpha_FeH}.  }
        \label{fig:Al_FeH}
\end{figure*}

In this work, we determine Al abundances of a large group of Scl stars for the first time.  There is a concern that NLTE correction for Al could be substantial ($\sim 0.4$ dex) for very metal-poor stars ([Fe/H$]<-2.0$) when using optical lines \citep{Andrievsky2008,Mashonkina2017}. However, \citet{Lind2022} suggested that the NLTE correction for Al lines in the APOGEE spectral range ($\lambda \sim 16719,16751,16763$ \AA) is negligible ($<0.1$ dex). \,\,\,\,
[Al/Fe] of Scl stars show a decreasing trend as the metallicity increases. This is expected considering that Al is mainly synthesized in massive stars.
Compared to the majority of MW field stars, [Al/Fe] of dwarf galaxies (Scl and Sgr) are systematically lower (Fig. \ref{fig:Al_FeH}, upper panel).  The deficient [Al/Fe] in Sgr was pointed out by \citet{Hasselquist2017}. 
Given the similar nucleosynthetic origins of Al and Mg, their ratio could be informative. In terms of [Al/Mg] (Fig. \ref{fig:Al_FeH}, lower panel), Scl stars show a flat trend as a function of metallicity, while Sgr and MW stars show clear increasing trends. This trend reflects the increasing yield of Al over $\alpha$-elements with initial stellar metallicity.
There are also a group of halo field stars with [Fe/H$]<-2.5$ showing [Al/Fe$]\sim -0.7$ and [Al/Mg]$\sim-1.0$. It has been suggested that these stars originate from dissolved dwarf galaxies \citep[e.g.,][]{Das2020}. Al is an excellent tracer with which to chemically tag stars of extragalactic origin.

\subsection{Iron-peak elements}

Though SNe Ia runaway deflagration obliterations of white dwarfs have a signature more tilted towards the iron-peak group \citep{Nomoto1997}, the solar composition of the iron-peak elements are in fact a heterogeneous combination of both SNe Ia and core-collapse SNe II \citep{Woosley1995}. As dwarf galaxies and MW have different star-formation timescales, this discrepancy may manifest itself in iron-peak elements. 

\begin{figure*}
        \centering \includegraphics[width=0.8\linewidth,angle=0]{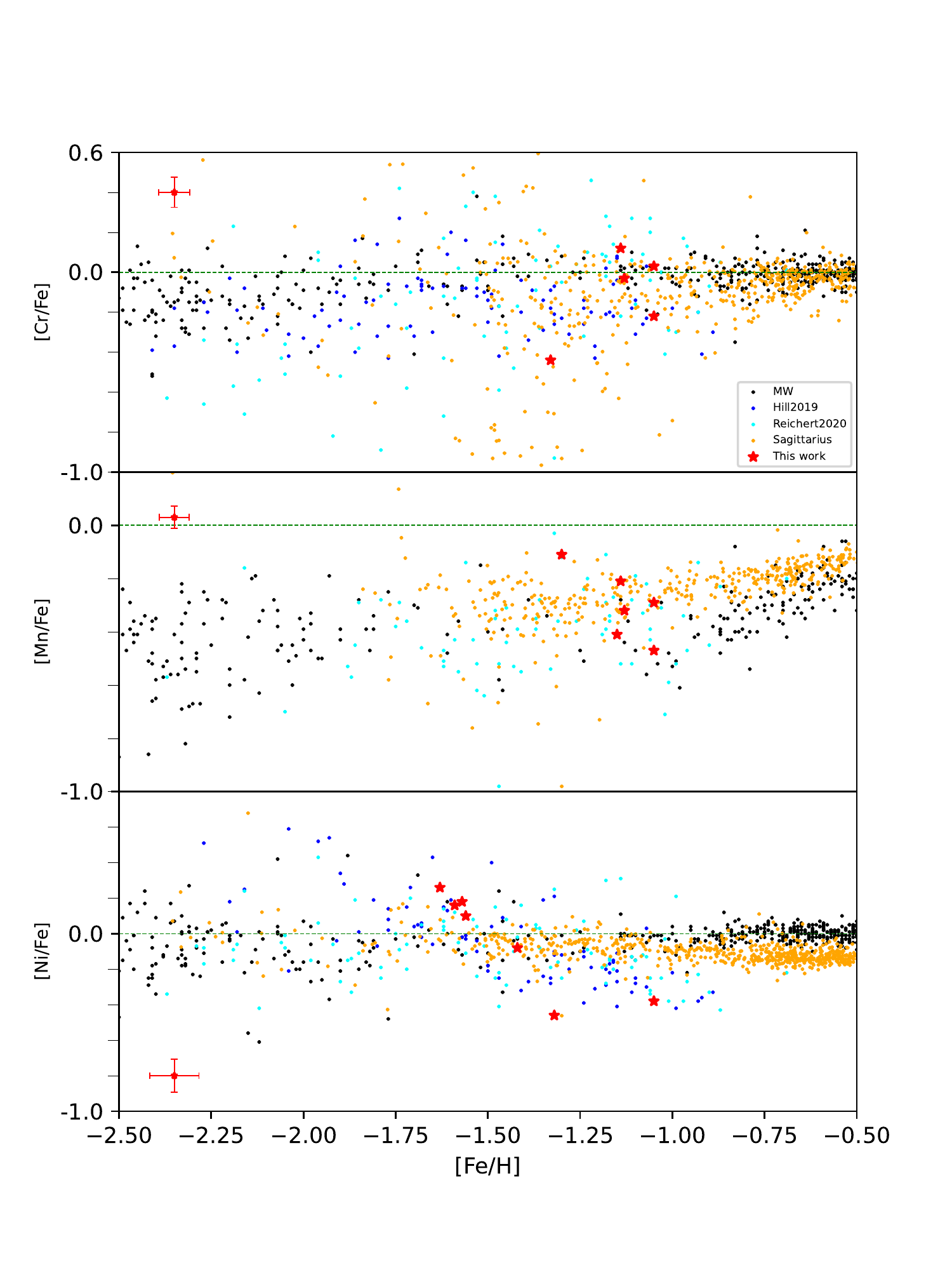}
        \caption{Abundances of iron peak elements vs. [Fe/H].  Symbols are the same as in Fig. \ref{fig:alpha_FeH}.}
        \label{fig:IronPeak_FeH}
\end{figure*}

As the iron-peak elements investigated in this study show weak atomic lines in the observed spectra, only a few measurements are available, mostly for more metal-rich stars.
 The upper panel of Fig.~\ref{fig:IronPeak_FeH} shows that our [Cr/Fe] abundances are generally consistent with those of \citetalias{Hill2019} and \citetalias{Reichert2020} in a similar metallicity range. Similar conclusions can be drawn for [Mn/Fe]. As our measurements of Cr and Mn are concentrated in a small metallicity range, to avoid over-interpreting our data, we refer to the aforementioned studies for their astrophysical implications.

Albeit based on a  small sample size, [Ni/Fe] shows a seemingly decreasing trend as a function of [Fe/H], which is similar to other $\alpha$-elements. This is consistent with the Ni trend observed by \citetalias{Hill2019}, who suggested this could be a result of the low contribution of Ni from SNe Ia (also see \citealt{Nissen2011}). However, the decreasing trend of Ni in \citetalias{Reichert2020} is less evident, indicating a more dedicated investigation of Ni is necessary.

\subsection{Neutron-capture elements}

Elements heavier than iron are mostly generated via neutron-capture processes. Depending on the relative speed of neutron capture compared to $\beta$-decay, the neutron-capture processes are divided into two types, rapid ({\it r}-) and slow ({\it s}-).  The astrophysical sites of {\it r}-process element production have been debated over the past 60 years \citep[e.g.,][]{Thielemann2011, Kajino2019}. The more popular models include core collapse SNe \citep[e.g.,][]{Woosley1994} and neutron star mergers \citep[e.g.,][]{Cote2018, Watson2019}. The main {\it s}-process elements are synthesized by AGB stars during thermal pulsations \citep{Busso2001, Karaks2014}. Two nuclear reactions are the major neutron excess sources in AGB stars: $\rm {^{13}C(\alpha,n) ^{16}O}$ and $\rm {^{22}Ne(\alpha,n) ^{25}Mg}$. The former reaction dominates the low-mass AGB stars, while the latter is mainly found in massive AGB stars \citep{Cristallo2015}. As the number of free neutrons per iron seed increases, the {\it s}-process flow first seeds the light {\it s}-process peak (Sr$-$Y$-$Zr), extending to $\rm ^{136}Ba$, and then reaches the heavy {\it s}-process peak (Ba$-$La$-$Ce$-$Pr$-$Nd), extending to $\rm ^{204}Pb-^{207}Pb$ \citep{Bisterzo2014}.

\begin{figure}
        \centering \includegraphics[width=\linewidth,angle=0]{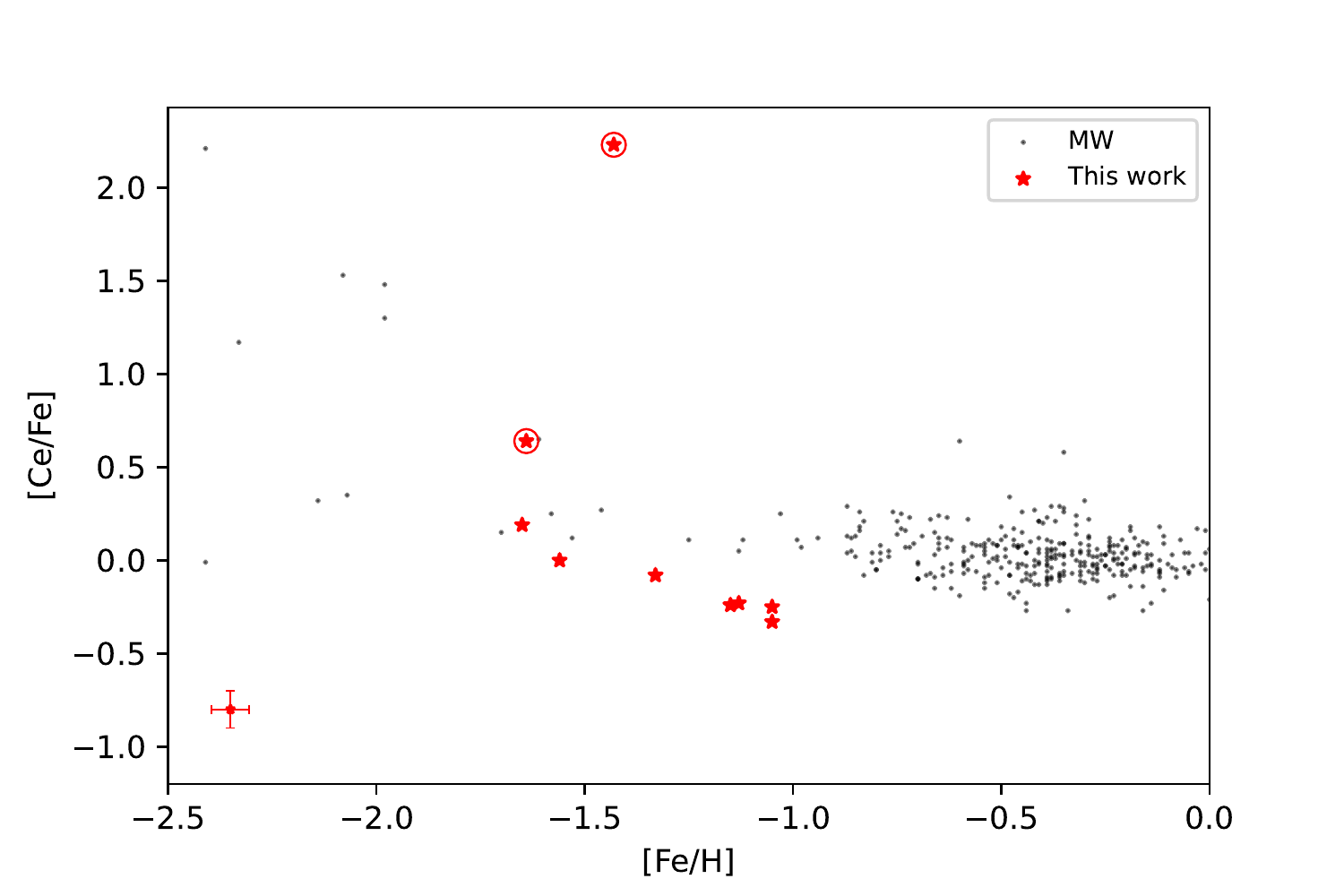}
        \caption{ [Ce/Fe] vs. [Fe/H]. Symbols are the same as in Fig.~\ref{fig:alpha_FeH}.}
        \label{fig:Ce}
\end{figure}

\begin{figure}
        \centering \includegraphics[width=\linewidth,angle=0]{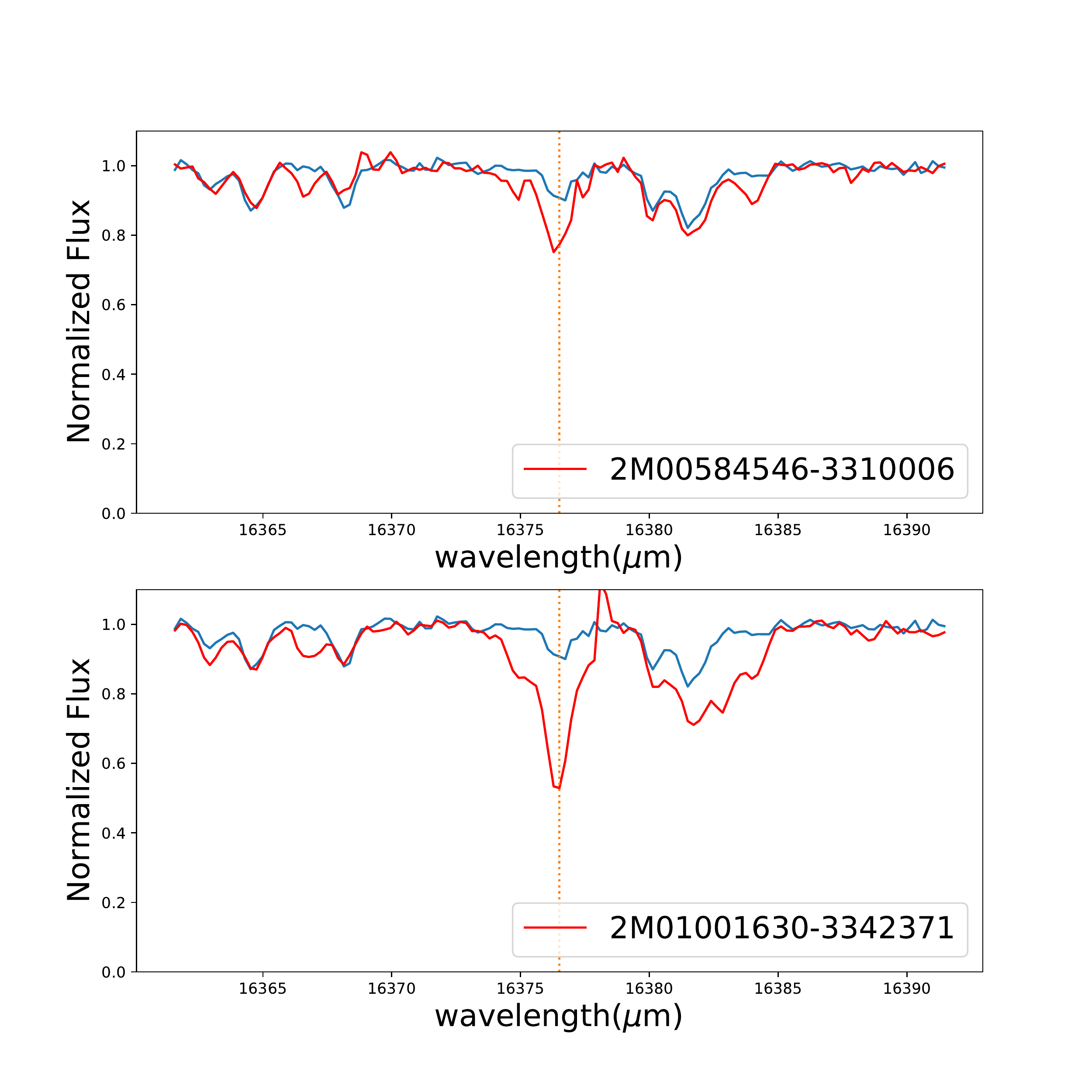}
        \caption{Spectra of two Ce-strong stars (red lines).  (T$_{\rm eff}$, $\log g$, [Fe/H]) = (4023.17 K, 0.5, -1.64) and (3959.43 K, 0.38, -1.43) for 2M00584546-3310006 and 2M01001630-3342371, respectively.  A star (2M01001215-3337260) with similar stellar parameters (4107.19 K, 0.50, -1.65) is also plotted (blue lines) for comparison. The orange dotted lines mark the center of the Ce feature. }
        \label{fig:Cespec}
\end{figure}

Nucleosynthetic models indicate that Ce is mostly produced by the {\it s}-process inside AGB stars \citep{Kobayashi2020}.  \citet{Bisterzo2014} estimated that 83\% of the Ce in our Sun is produced by the {\it s}-process, which is comparable to other typical {\it s}-process-dominated elements; for example, Ba (85\%).
The Ce abundances of our Scl stars are compared with MW stars in Fig.~\ref{fig:Ce}. Two stars are found to have exceptionally large [Ce/Fe], where strong Ce lines in the spectra (Fig. \ref{fig:Cespec}) support the abundance credibility. More interestingly, these stars also have large [C/Fe] and [N/Fe] abundances; we further discuss these in Sect. \ref{subsec:Ce}.  After excluding the two stars with the strongest [Ce/Fe], a decreasing trend as a function of metallicity seems to exist. However, this trend is rather uncertain because of the low number statistics and the weakness of Ce lines. On the other hand, another typical {\it s}-process element, Ba, shows an increasing trend as [Fe/H] increases for Scl stars \citepalias{Hill2019}; Ce in other dwarf galaxies with larger galaxy mass than Scl also shows an increasing trend as a function 
of metallicity \citep{Hasselquist2021}. Further high-S/N spectra with measurable Ce lines are needed to confirm this trend.

\section{Discussion}
\label{sec:discussion}

\subsection{Chemical evolution model}

 Scl has been studied extensively using different galaxy chemical evolution (GCE) models. Here, we test whether our new NIR measurements of different element abundances are consistent with the expected element abundance ratios given by GCE calculations. In particular, we calculate the evolution of [Al/Fe] and [O/Fe] values of Scl and compare them with observations  for the first time.
We used the publicly available GCE code ``GalIMF'' \citep{2017A&A...607A.126Y,2019A&A...629A..93Y} to generate possible evolution tracks of element abundance ratios. The code adopts an empirically calibrated environment-dependent IMF variation law (the IGIMF theory, \citealt{2013pss5.book..115K,2018A&A...620A..39J,2020A&A...637A..68Y}), which becomes top light in galaxies with a low star formation rate (SFR; \citealt{2009ApJ...706..599L,2020A&A...637A..68Y,2021NatAs...5.1247M}). The SN Ia rate in the GalIMF code is a function of the IMF because the number of potential SN Ia progenitor stars depends on it \citep{2019A&A...629A..93Y,2021A&A...655A..19Y}. We adopt the yield tables in \citet{2010MNRAS.403.1413K}, \citet{2018ApJS..237...13L}, and \citet{1999ApJS..125..439I} for AGB, SN II, and SN Ia, respectively. More details of our GCE model are provided in Appendix~\ref{Appendix:GCE model}.

We assume a delayed-$\tau$ star formation history (Eq.~\ref{eq:SFR}) that is similar to previous GCE studies of Scl (\citealt{2014MNRAS.441.2815V, 2022ApJ...925...66D}, see Fig.~\ref{fig:SFH}). The resulting time-integrated galaxy-wide IMF has a top-light shape similar to the power-law IMF with a power index for massive stars of about $-2.7$ (see Section~\ref{sec: IMF variation}). Different SN Ia delay times ($t_{\rm delay, min}=40$, 100, and 400~Myr in Eq.~\ref{eq:DTD}) are tested and $t_{\rm delay, min}=100$~Myr is preferred, in agreement with \citet{2022ApJ...925...66D}. 
The present-day stellar mass, $2.61\cdot 10^6~M_\odot$, and the mean stellar metallicity, ${\rm [Fe/H]}=-1.67$, of our best-fit model fit well with the literature values (see Section~\ref{sec:SFH} and \citealt{2011ApJ...727...78K}).

\begin{figure}[!hbt]
    \centering
    \includegraphics[width=9cm]{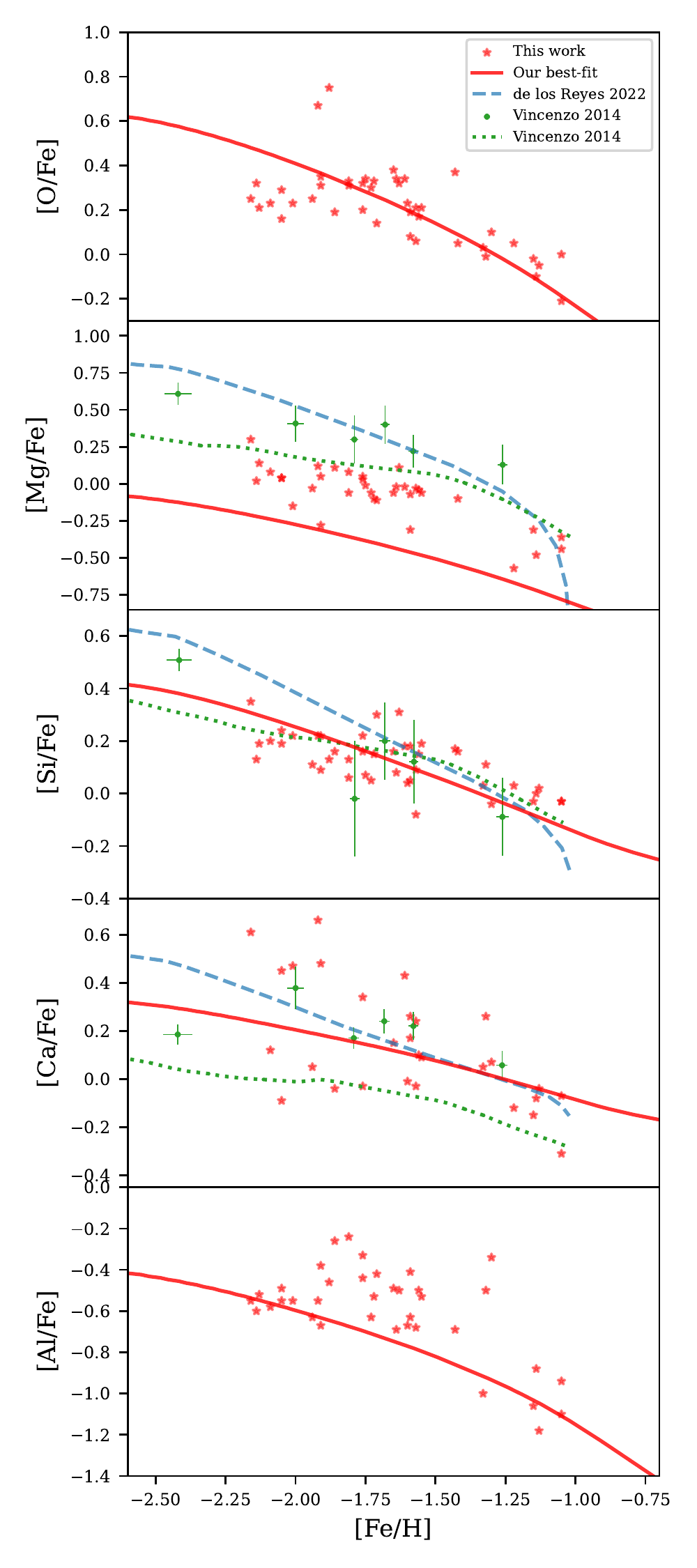}
    \caption{Evolution of [X/Fe]--[Fe/H] relations of our best-fit GCE model (red curves) in comparison with observations (red stars, Table~\ref{tab:abundances}).
    The green dots are data compiled in \citet{2014MNRAS.441.2815V}. The green dotted curves and blue dashed curves are the dotted-line model in \citet[their figure 3,]{2014MNRAS.441.2815V} and the fiducial model in \citet{2022ApJ...925...66D}, respectively.}
    \label{fig:abundances}
\end{figure}

The best-fit evolution tracks of element abundance ratios ([O/Fe], [Mg/Fe], [Si/Fe], [Ca/Fe], and [Al/Fe]) are demonstrated (Fig.~\ref{fig:abundances}) together with our measurement values listed in Table~\ref{tab:abundances}. Results of previous GCE studies of Scl are shown for comparison (the dotted-line model in \citealt[their figure 3]{2014MNRAS.441.2815V} and the fiducial model in \citealt{2022ApJ...925...66D}). The ``knee'' of the evolution tracks at about ${\rm [Fe/H]}\approx-1.5$ is caused by an increased ratio between SN~Ia and SN~II after the peak SFR, but this feature is not prominent (cf. \citealt{2014MNRAS.441.2815V} and \citealt{2022ApJ...925...66D}). The evolution of [Ti/Fe], [Cr/Fe], [Mn/Fe], [Ni/Fe], and [Ce/Fe] is not shown because these have a larger observational uncertainty or stellar yield uncertainty.

The measured evolution of [O/Fe], [Si/Fe], and [Ca/Fe] shows good agreement with our GCE model (Fig.~\ref{fig:abundances}).
Similar to previous studies \citep{2004MNRAS.351.1338L,2006ApJ...646..184F,2013MNRAS.434..471R,2017ApJ...835..128C}, the [Mg/Fe] value from our best-fit model appears to be about 0.3~dex lower than the observed values, which cannot be improved without applying certain modifications of the adopted yield table (c.f. \citealt{2022ApJ...925...66D}). The Al yield appears to face a similar problem, although not as significant as that found for the Mg yield. 
The systematic difference between the measured and calculated evolution of [Mg/Fe] and [Al/Fe] suggests problems in either the adopted stellar element yield tables or the measurements, which requires further investigation in the future.

\subsection{Radial chemical gradients}
\label{subsec:cg}

\begin{figure*}
        \centering \includegraphics[width=1.0\linewidth,angle=0]{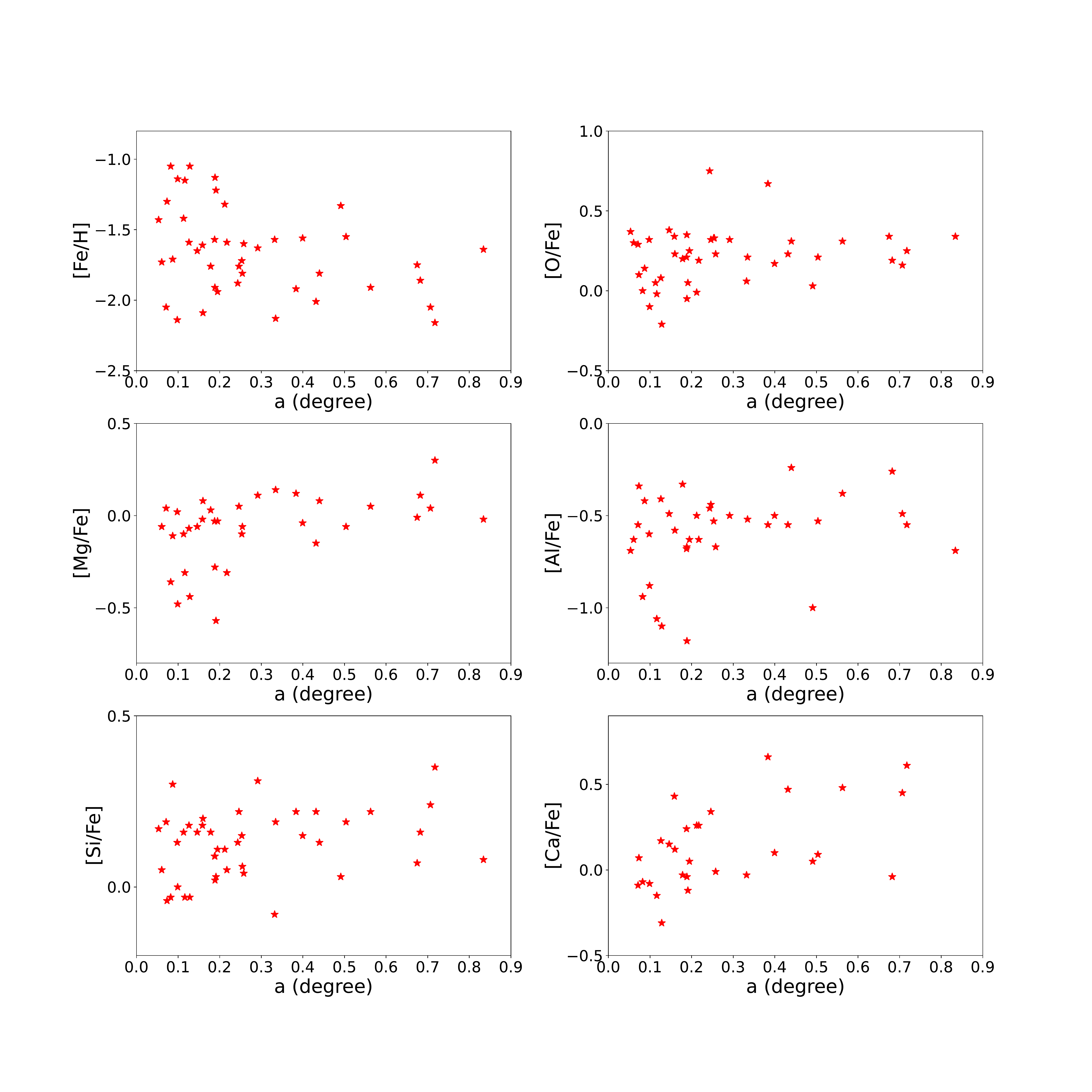}
        \caption{Element abundances as a function of elliptical radius in degrees. 
        }
        \label{fig:abu_radial}
\end{figure*}

\begin{table*}
        \caption{Results of linear least squares regression for each panel in Fig. \ref{fig:abu_radial}}
        \label{tab:XFe_a_linearFitting_coefficient}
        \begin{center}
\begin{tabular}{c c c c c}
\hline
\hline
\hline
element & k & k error & r-value & p-value \\
    & (dex/degree) &(dex/degree) & &  \\
\hline
\hline
  {\rm [Fe/H]} & -0.536 & 0.21859 & -0.358 & 0.019\\
  {\rm [O/Fe]} & 0.175 & 0.13338 & 0.201 & 0.197\\
  {\rm [Mg/Fe]} & 0.399 & 0.13371 & 0.461 & 0.005\\
  {\rm [Al/Fe]} & 0.261 & 0.17559 & 0.244 & 0.145\\
  {\rm [Si/Fe]} & 0.138 & 0.07194 & 0.286 & 0.063\\
  {\rm [Ca/Fe]} & 0.622 & 0.2038 & 0.506 & 0.005\\
\hline
\hline
\end{tabular}
        \end{center}
        \raggedright{\hspace{2cm}Notes: k: best-fit slope . k error: error of the best-fit slope. r-value: linear correlation coefficient.\\ \hspace{2cm}p-value: two-sided p-value for a hypothesis test in which the null hypothesis is that the slope is zero.\\}
\end{table*}

\citet{Bettinelli2019} suggested that the star formation history of Scl may change radially: the innermost region is expected to have undergone a longer period of star formation ($\sim$ 1.5 Gyr) compared to the outermost region ($\sim$ 0.5 Gyr). Given that star formation strongly influences chemical abundances, chemical gradients are helpful to disentangle the detailed evolution of galaxies. For example, using open clusters and young field stars in the Galactic disc, \citet{Magrini2017} confirmed the negative radial gradient of [Fe/H], and discovered positive radial gradients for [O/Fe], [Si/Fe], and [Ca/Fe], and a flat radial gradient for [Mg/Fe]. 

In this work, we discuss the radial chemical gradients of Scl for the first time thanks to the wide spatial coverage of our sample stars. We plot [Fe/H], [O/Fe], [Mg/Fe], [Al/Fe], [Si/Fe], and [Ca/Fe] as a function of  elliptical radius in Fig. \ref{fig:abu_radial}. Other elements are not discussed here because of insufficient data. 
 To explore the correlations between abundances and elliptical radius, we preformed linear least squares regression for each panel, and we list their slopes (k), slope errors, r-values (correlation coefficients), and p-value\footnote{A hypothesis test, whereby the null hypothesis is that the slope is zero, using Wald Test with t-distribution of the test statistic.} in Table \ref{tab:XFe_a_linearFitting_coefficient}. \citet{Tolstoy2004} found a negative radial metallicity gradient in Scl using the Ca II triplet. Interestingly, there is a visible slope change around 0.2$^{\circ}$ in elliptical radius (their Fig. 3). Our results agree with those of \citet{Tolstoy2004}: the p-value$= 0.019$ and $k=-0.536$ that we find suggest a negative radial metallicity gradient, and there is a hint that the slope becomes flatter beyond 0.2$^{\circ}$ elliptical radius. 
 On the other hand, our linear regression analysis (Table \ref{tab:XFe_a_linearFitting_coefficient}) suggests a nondetectable radial gradient of [O/Fe], [Al/Fe], and [Si/Fe], as their two-sided p-value is too large. Meanwhile, [Mg/Fe] and [Ca/Fe] show positive radial gradients (positive slope values), albeit with large scatter ($0.4 <$ r-value $\lesssim 0.5$). The two-sided p-values are less than 0.05, indicating reliable correlations. 
The radial metallicity gradient and radial gradients of [Mg/Fe] and [Ca/Fe] can be explained by the outside-in scenario proposed by \citet{Bettinelli2019} and also found in other dwarf galaxies in the Local Group \citep{Hidalgo2013}. The prolonged star formation in the innermost region ($\sim$ 1.5 Gyr, \citealt{Bettinelli2019}) generates comparatively more iron, causing higher [Fe/H], and comparatively lower [Mg/Fe] and [Ca/Fe]. A chemical evolution model with radial distribution will be the helpful to quantify these.

\subsection{Stars with enhanced Ce}
\label{subsec:Ce}

\begin{figure*}
        \centering \includegraphics[width=0.7\linewidth,angle=0]{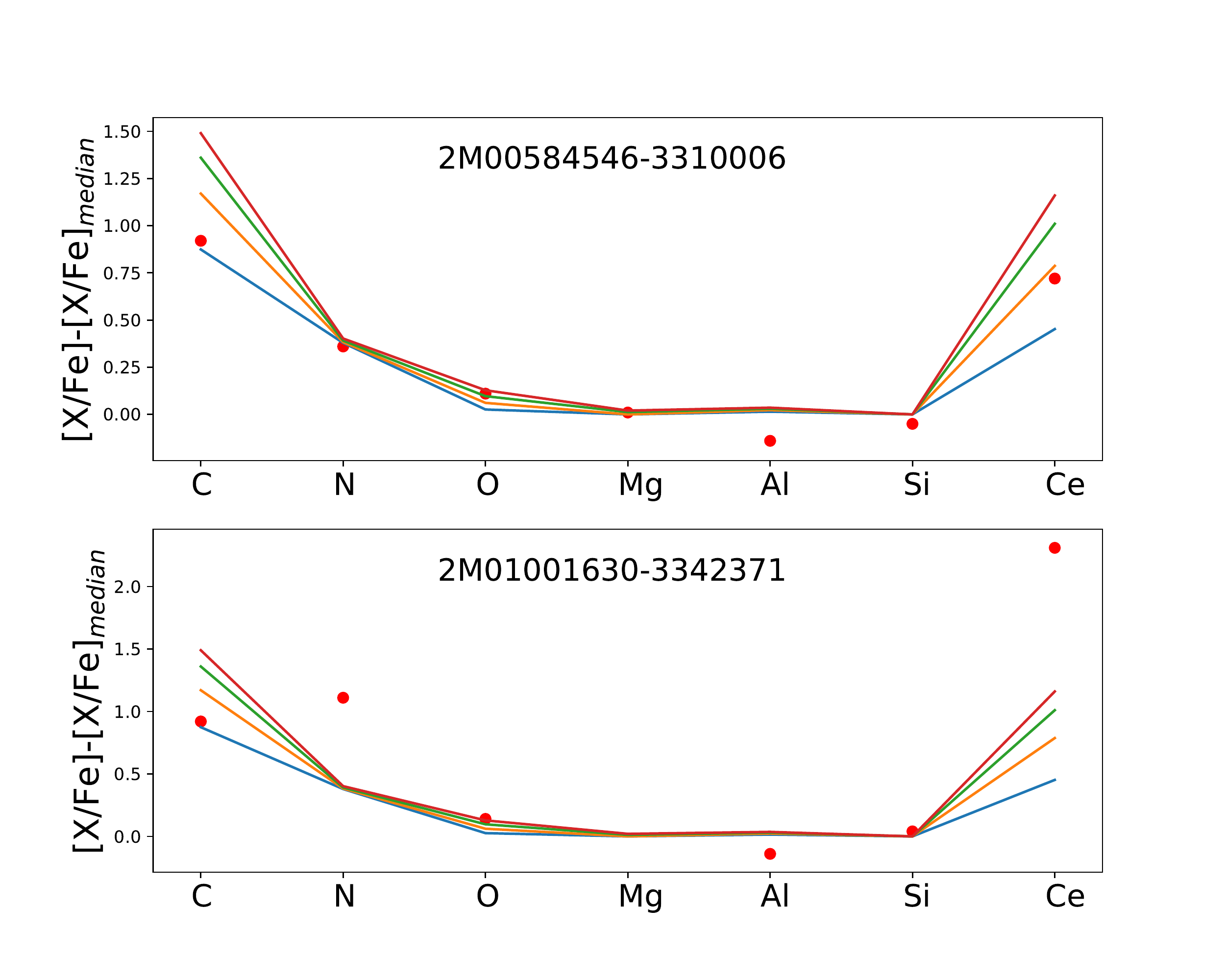}
        \caption{Chemical abundances of two Ce-enhanced stars (red dots) compared with the AGB model of 1.3 $M_{\odot}$, Z=0.0001. Models of second, third, fourth, and fifth episodes of third dredge-up are shown as blue, orange, green, and red lines, respectively. }
        \label{fig:AGB}
\end{figure*}

As we discuss in Section \ref{sec:result}, two stars show enhanced C, N, and Ce, which is a strong hint of AGB stars. Their chemical abundances are compared to those suggested by the AGB model of 1.3 $M_{\odot}$, Z=0.0001\footnote{http://fruity.oa-teramo.inaf.it/modelli.pl} \citep{Cristallo2011}. To minimize the impact of chemical abundances inherited from the original cloud, the median of a given element ([X/Fe]$_{median}$) is subtracted from the measured abundance\footnote{ Because N abundances are only measured in three stars, and they do not reflect the original abundances of the cloud;  we therefore use [X/Fe]$_{median} = 0.5$ instead.} (Fig.~\ref{fig:AGB}). AGB stars going through third dredge-up (TDU) experience multiple episodes, where chemical composition changes accordingly. A good match between observations and models is seen in  2M00584546-3310006, while 2M01001630-3342371 detects higher N and Ce abundances than suggested by AGB models. More massive models or models with different metallicities do not seem to fully reconcile the discrepancies. AGB models of $<1.3 M_{\odot}$ or binary models may be needed to explain such high N and Ce abundances. 

\subsection{N-rich field stars}
\label{subsec:Nrich}

The recently found N-rich field stars in the MW \citep{martell2010,fernandez_trincado2017,Tang2019,fernandez_trincado2019,Tang2020} have fuelled debate  as to their origin. It is suggested that most N-rich field stars may be formed in the dense environment of globular clusters \citep{martell2010,Tang2020}, while some of them are generated in binary systems \citep{FTbinary}. According to chemo-dynamical analysis of N-rich field stars, a substantial portion of them may have an extragalactic origin \citep{Tang2020,Yu2021,FT2022}. \citet{FTSgr,FTMC} reported the discovery of a large group of N-rich field stars in Sgr and Magellanic Clouds. 

In this work, we find most of the $^{12}$C$^{14}$N molecular lines are too weak to give meaningful results, except for three stars with exceptionally large [N/Fe]. As mentioned in Section \ref{subsec:Ce}, two stars with strong C, N, and Ce are likely AGB stars. While the third star, 2M00591209-3346208, does show large [N/Fe] ($\sim 1.4$), its [C/Fe] is also quite large ($\sim 0.6$). This star therefore fails to satisfy one of the criteria of N-rich field stars ([C/Fe]$<0.15$, \citealt{schiavon2017,fernandez_trincado2019}). Given that the nondetection of strong $^{12}$C$^{14}$N molecular lines in other Scl stars excludes the possibility of them being N-rich field stars, we conclude that no N-rich field stars are detected in our Scl sample. Our result is consistent with that of \citet{lardo2016}, who used optical CN and CH bands around 4000 \AA~to investigate the C and N abundances of Scl stars. The lack of N-rich field stars in Scl may indicate that this dwarf galaxy is deficient of very dense star forming environments (e.g., globular clusters).

\section{Conclusions}
\label{sec:conclusion}

Dissolved and existing dwarf galaxies are the basic building blocks of our past, present, and future MW. The detailed evolutionary history of galaxy formation is imprinted in chemical abundances, waiting to be decoded. In this work, we analyze C, N, O, Mg, Al, Si, Ca, Ti, Cr, Fe, Mn, Ni, and Ce of 43 giant stars in Scl using NIR APOGEE spectra. Precise O, Si, and Al abundances are determined in an unprecedented sample size of stars ($\sim$40), thanks to the strong feature lines in APOGEE spectra.
After comparing with abundances obtained from optical spectra, there is generally good agreement, except for [Mg/Fe]. [$\alpha$/Fe] of Scl stars is systematically lower than in MW field stars or Sgr stars, confirming the observed trend that less massive galaxies show lower [$\alpha$/Fe].  
{[Al/Fe] shows a decreasing trend as a function of metallicity, with a median value of about -0.5, providing an excellent indicator with which to distinguish Scl stars from the field stars of the MW. }

 The observed abundance ratios are compared with the calculated chemical evolution of Scl assuming an environment-dependent IMF given by IGIMF theory.
Similar to previous GCE studies, our model favors a top-light IMF with a high-mass IMF power index of about $-2.7$ and a minimum SN~Ia delay time of about 100~Myr. The expected evolution tracks of [O/Fe], [Si/Fe], and [Ca/Fe] from the GCE model agree well with our observation. With the Al abundances of a large number of stars in Scl measured for the first time, we find that the GCE-predicted [Al/Fe] evolution track is consistent with observations. However, the observed [Al/Fe] values are generally higher than those predicted by the GCE.

 Our linear regression analysis suggests a negative radial metallicity gradient in Scl, spanning up to 0.9 degrees in elliptical radius. [Mg/Fe] and [Ca/Fe] show increasing trends as a function of metallicity. These are consistent with the inside-out formation scenario, which suggests the
quenching of star formation toward the center as the galaxy runs out of gas in the outskirts. Furthermore, we detect no N-rich field stars in our sample, indicating that Scl may be deficient in very dense star forming environment. In the near future, the China Space Station Telescope (CSST), equipped with the most advanced ultraviolet and optical photometric and spectroscopic system, will have a significant impact on the investigation of the star formation history across dwarf galaxies, the search for N-rich field stars and AGB stars, and so on.

\section*{Acknowledgments}
We thank Thomas Masseron, Gary Da Costa, Yang Huang for helpful discussions. 
We thank the anonymous referee for insightful comments.
B.T. and J.Z. gratefully acknowledge support from the National Natural Science Foundation of China under grant No. U1931102, the China Manned Space Project NO. CMS-CSST-2021-B03, the Natural Science Foundation of Guangdong Province under grant No. 2022A1515010732, the National Natural Science Foundation of China through grants No. 12233013, and the Fundamental Research Funds for the Central Universities, Sun Yat-sen University under No. 22qntd3101.
Z.Y. acknowledges support from the Fundamental Research Funds for the Central Universities under grant number 0201/14380049, support through the Jiangsu Funding Program for Excellent Postdoctoral Talent under grant number 2022ZB54, and support of the National Natural Science Foundation of China under grant No. 12203021.
Z.Z. and Z.Y. acknowledge the support of the National Natural Science Foundation of China under grants No. 12041305, 12173016, the Program for Innovative Talents, Entrepreneurs in Jiangsu, the science research grants from the China Manned Space Project with NO.CMS-CSST-2021-A08 (IMF), and the science research grants from the China Manned Space Project with NO.CMS-CSST-2021-A07 (lensing).
The development of our chemical evolution model applied in this work (GalIMF) benefited from the International Space Science Institute (ISSI/ISSI-BJ) in Bern and Beijing, thanks to the funding of the team “Chemical abundances in the ISM: the litmus test of stellar IMF variations in galaxies across cosmic time” (Donatella Romano and Zhi-Yu Zhang).
J.G.F-T gratefully acknowledges the grant support provided by Proyecto Fondecyt Iniciaci\'on No. 11220340, and also from ANID Concurso de Fomento a la Vinculaci\'on Internacional para Instituciones de Investigaci\'on Regionales (Modalidad corta duraci\'on) Proyecto No. FOVI210020, and from the Joint Committee ESO-Government of Chile 2021 (ORP 023/2021), and from Becas Santander Movilidad Internacional Profesores 2022, Banco Santander Chile.

Funding for the Sloan Digital Sky Survey IV has been provided by the Alfred P. Sloan Foundation, the U.S. Department of Energy Office of Science, and the Participating Institutions. SDSS- IV acknowledges support and resources from the Center for High-Performance Computing at the University of Utah. The SDSS website is http://www.sdss.org.

SDSS-IV is managed by the Astrophysical Research Consortium for the Participating Institutions of the SDSS Collaboration including the Brazilian Participation Group, the Carnegie Institution for Science, Carnegie Mellon University, the Chilean Participation Group, the French Participation Group, Harvard-Smithsonian Center for Astrophysics, Instituto de Astrof\`{i}sica de Canarias, The Johns Hopkins University, Kavli Institute for the Physics and Mathematics of the Universe (IPMU) / University of Tokyo, Lawrence Berkeley National Laboratory, Leibniz Institut f\"{u}r Astrophysik Potsdam (AIP), Max-Planck-Institut f\"{u}r Astronomie (MPIA Heidelberg), Max-Planck-Institut f\"{u}r Astrophysik (MPA Garching), Max-Planck-Institut f\"{u}r Extraterrestrische Physik (MPE), National Astronomical Observatory of China, New Mexico State University, New York University, University of Notre Dame, Observat\'{o}rio Nacional / MCTI, The Ohio State University, Pennsylvania State University, Shanghai Astronomical Observatory, United Kingdom Participation Group, Universidad Nacional Aut\'{o}noma de M\'{e}xico, University of Arizona, University of Colorado Boulder, University of Oxford, University of Portsmouth, University of Utah, University of Virginia, University of Washington, University of Wisconsin, Vanderbilt University, and Yale University.

\bibliographystyle{aa}
\bibliography{nrich,Sclref,survey,library}

\begin{appendix}

\section{GCE model}\label{Appendix:GCE model}
\subsection{Method}

Metal-poor gas-rich dwarf galaxies are observed to be dispersion-dominated systems with high specific SFRs \citep{2022arXiv220604709I}, in agreement with the idea that spheroids are formed in a quick monolithic collapse \citep{2009A&A...499..409C}.
Under this framework, we implement a closed-box model, as described by \citet{2020A&A...637A..68Y}, to calculate the chemical evolution of Scl. 
We assume that a primordial gas of $1.26\cdot 10^8~M_\odot$ (similar to \citealt{2014MNRAS.441.2815V}) is presented at the beginning of the simulation while there is no additional gas infall. 
Stars form according to a given star formation history. The SFR for each 10 Myr simulation time step, $\bar{\psi}_{\rm 10~Myr}$, is given by the delayed-$\tau$ model,
\begin{equation}\label{eq:SFR}
    \bar{\psi}_{\rm 10~Myr}(t) = R \cdot t/\tau \cdot e^{-t/\tau},
\end{equation}
where $R$ and $\tau$ are characteristic SFR and star formation timescale constants (best-fit parameter values given in Section~\ref{sec:SFH}).

The AGB and core-collapse supernovae yields are adopted from \citet{2010MNRAS.403.1413K} and \citet{2018ApJS..237...13L}, respectively, from the readily complied yield table provided by the NuPyCEE code \citep{2018ApJS..237...42R}. 
The IMF-weighted yields of explosive $\alpha$ elements (Si and Ca) is more pronounced for stars between 8 and $12~M_\odot$ in \citet{2018ApJS..237...13L}, considering stellar rotation, relative to \citet{2013ARA&A..51..457N}. In general, this leads to a better fit with observed [X/Fe]--[Fe/H] relations in dwarf galaxies (cf. \citealt{2021ApJ...910..114M}).
We find that a mixture of core-collapse supernova yields (averaging the yields for nonrotating stars and the yields for stars with an initial equatorial velocity of 150~km~s$^{-1}$ in \citealt{2018ApJS..237...13L}) lead to the best fit to the observed abundance ratio evolution. 

For the SN~Ia yield, we adopt \citet[their W70 model]{1999ApJS..125..439I}. The delay time distribution (DTD) of SN~Ia, $f_{\rm delay}(t)$, is assumed to be a power law \citep{2012PASA...29..447M} with a variable minimum SN~Ia delay time, $t_{\rm delay, min}$,
\begin{equation}\label{eq:DTD}
    f_{\rm delay}(t) =
    \begin{cases}
        0, & t \leqslant t_{\rm delay, min}, \\
        k~{\rm [yr}^{-1}M_\odot^{-1}{\rm ]} \cdot (t/1~{\rm Gyr})^{-1}, & t>t_{\rm delay, min}.
    \end{cases}
\end{equation}

In the above equations, the normalization parameter $k$ determines the total number of SNe~Ia, which depends on the number of available SN~Ia progenitors, and therefore on the IMF.
The number of SN~Ia per stellar mass formed within time $t$ after the formation of a simple stellar population, $n_{\rm Ia}$, is scaled according to the IMF variation, following \citet{2019A&A...629A..93Y,2021A&A...655A..19Y}. That is,
\begin{equation}\label{eq:SNIa rate}
    n_{\rm Ia}(t, \xi)=
    \frac{n_{1.5,m_{\rm MWD}}(\xi)}{M_{0.08,150}(\xi)} 
    \cdot \frac{n_{1.5,m_{\rm MWD}}(\xi)}{n_{0.08,150}(\xi)} 
    \cdot \int_{0}^tf_{\rm delay}(t)\mathrm{d}t,
\end{equation}
where $\xi$ is the IMF of the stellar population, $m_{\rm MWD}$ is the mass of the most massive star that can become a SN~Ia progenitor, $n$ is the number of stars within the mass range given by its subscript (in the unit of $M_\odot$), and $M$ is the mass of stars in a given mass range indicated by its subscript. 

SN Ia delay times of $t_{\rm delay, min}=40$, 100, and 400~Myr are tested in the present study. As $t_{\rm delay, min}$ is the lifetime of the stars with mass $m_{\rm MWD}$ according to the stellar mass--lifetime relation (\citealt[their figure 3]{2019A&A...629A..93Y}), these two parameters are linked.
Therefore, $m_{\rm MWD}$ is set to 8, 5, and $3~M_\odot$ for these $t_{\rm delay, min}$ values respectively.
Then, the parameter $k$ with a given $t_{\rm delay, min}$ can be determined by the empirical SN~Ia production efficiency given by \citet{2012PASA...29..447M}, assuming the canonical \citet{2001MNRAS.322..231K} IMF, $\xi_{\mathrm{canonical}}$,
\begin{equation}\label{eq:normalization}
    n_{\rm Ia}(t=10~\mathrm{Gyr}, \xi_{\mathrm{canonical}})=0.002/M_\odot.
\end{equation}
See also \citet{2018MNRAS.479.3563F}, \citet{2021MNRAS.502.5882F}, and \citet{2021A&A...655A..19Y} for additional observational constraints on the SN~Ia production efficiency.
Therefore, $k$ is first calibrated for a given $t_{\rm delay, min}$ value, assuming the canonical IMF. 
It turns out that the $t_{\rm delay, min}=100$~Myr model fits best with the observed shape of [X/Fe]--[Fe/H] relations, which is consistent with the findings of \citet{2022ApJ...925...66D}.
For $t_{\rm delay, min}=100$~Myr, $k\approx 4.27\cdot 10^{-13}$~[yr$^{-1}M_\odot^{-1}$] (cf. \citealt[their equation~13]{2012PASA...29..447M}). Therefore, $n_{\rm Ia}(t)$ can be calculated for any variable IMF.
The DTD and therefore the exact values for the lower and upper mass limit for SN~Ia progenitors (1.5 and $m_{\rm MWD}$) are not well constrained by observations. 
Fortunately, the calculation result of Eq.~\ref{eq:SNIa rate} is not sensitive to these mass limits (cf. \citealt{2019A&A...629A..93Y,2021A&A...655A..19Y}).

\subsection{Best-fit result}

\subsubsection{Star formation history}\label{sec:SFH}

\begin{figure}[!hbt]
    \centering
    \includegraphics[width=9cm]{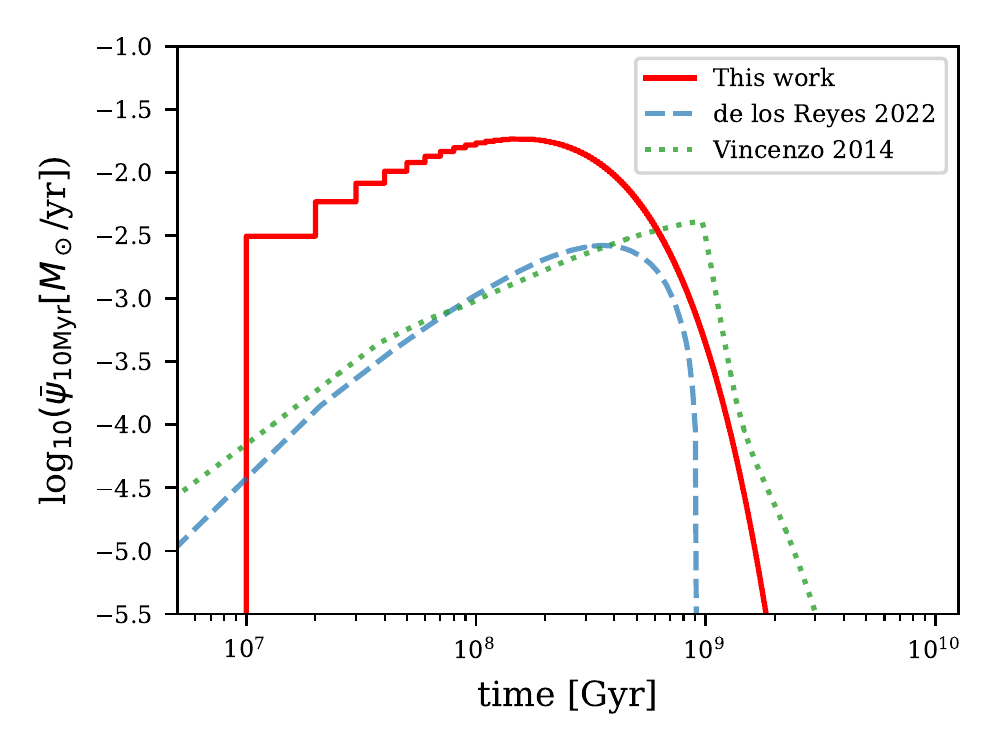}
    \caption{Star formation history of our best-fit model (red curve), formulated by the delayed-$\tau$ model (Eq.~\ref{eq:SFR}). The green dotted and blue dashed curves denote the star formation history applied in \citet{2014MNRAS.441.2815V} and \citet[their fiducial model]{2022ApJ...925...66D}, respectively.}
    \label{fig:SFH}
\end{figure}

The star formation history of our best-fit model is shown in Fig.~\ref{fig:SFH}. With the star formation history parameters (Eq.~\ref{eq:SFR}) being $R=0.05~M_\odot/$yr and $\tau=150$~Myr. The total stellar mass ever formed is $M_{\star}=\int_0^{\rm 13~Gyr}\bar{\psi}_{\rm 10~Myr}(t) {\rm d}t=R\cdot\tau=7.5\cdot 10^6~M_\odot$ and the resulting present-day stellar mass is $2.61\cdot 10^6~M_\odot$, which is comparable with observational values \citep{2008MNRAS.390.1453W,1998ARA&A..36..435M,2003AJ....125.1926G} and previous Scl models (\citealt{2014MNRAS.441.2815V} and \citealt{2022ApJ...925...66D}). 

\subsubsection{IMF variation}\label{sec: IMF variation}

The canonical multiple-part power-law stellar IMF has the form:
\begin{equation}\label{eq:xi}
\xi(m) =
\begin{cases} 
k_1 m^{-\alpha_1}, & 0.08 \leqslant m<0.5, \\ 
k_2 m^{-\alpha_2}, & 0.5 \leqslant m<1, \\
k_3 m^{-\alpha_3}, & 1 \leqslant m<150,
\end{cases}
\end{equation}
where $\alpha_1=1.3$ and $\alpha_2=\alpha_3=2.3$ \citep{2001MNRAS.322..231K,2002Sci...295...82K,2013pss5.book..115K}. The normalization parameters, $k_1$, $k_2$, and $k_3$ are adjusted to ensure a continuous IMF.

\begin{figure}[!hbt]
    \centering
    \includegraphics[width=9cm]{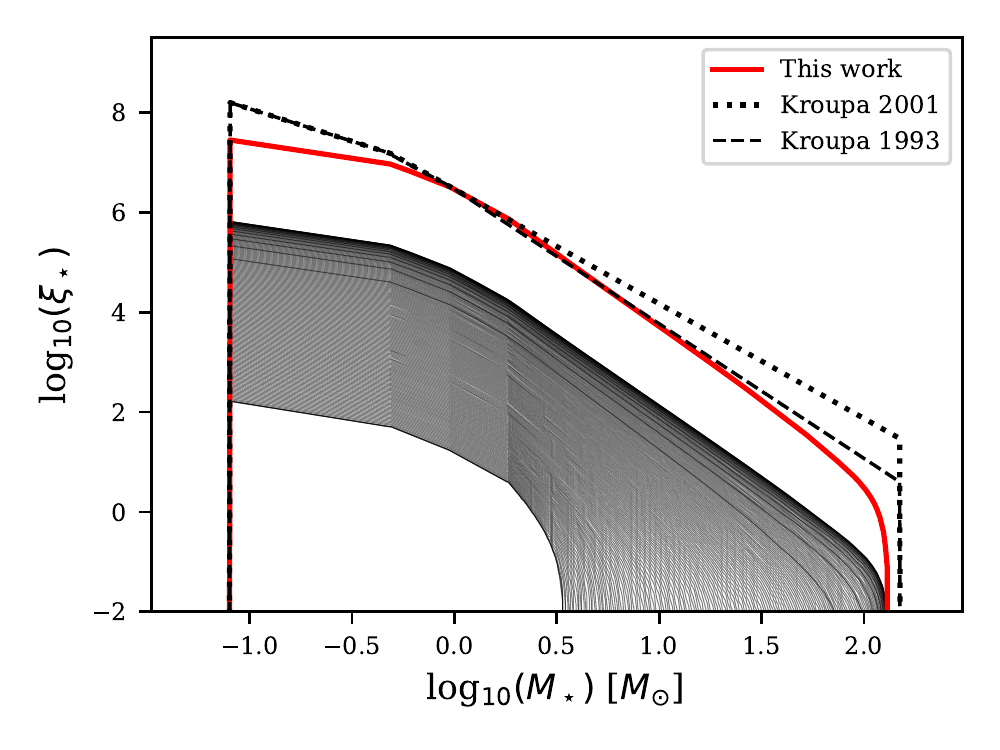}
    \caption{Time-integrated galaxy-wide IMF for all stars ever formed (thick solid red curve) and the galaxy-wide IMF for each 10~Myr star formation time step (thin solid curves) normalized by the total mass of a stellar population.
    Our best-fit IMF is compared with the canonical \citet{2001MNRAS.322..231K} IMF with $\alpha_3=2.3$ (dotted line) and the solar-neighborhood IMF from \citet{1993MNRAS.262..545K} with $\alpha_3=2.7$ (dashed line).
    }
    \label{fig:IMF}
\end{figure}

Figure~\ref{fig:IMF} shows the galaxy-wide IMFs at each time step and the time-integrated galaxy-wide IMF for the entire star formation history of our Scl model, calculated with the IGIMF formulation given in \citet{2020A&A...637A..68Y}. With a low rate of star formation (Fig.~\ref{fig:SFH}), Scl should have a top-light IMF (i.e., larger $\alpha_3$, cf. \citealt{2009ApJ...706..599L}). The IMF is also bottom-light (i.e., smaller $\alpha_1$ and $\alpha_2$) given the low stellar metallicities \citep{2013ApJ...771...29G}.

\end{appendix}

\end{document}